\documentclass[12pt]{iopart}
\usepackage{graphicx}
\usepackage{epstopdf, epsfig, rotating, floatpag, tikz}

\usepackage{subcaption}

\usepackage{bm}
\usepackage[hidelinks=true]{hyperref}
\usepackage{cleveref}
\usepackage{braket}
\usepackage{todonotes}
\usepackage{orcidlink}
\begin{document}

\title{Bifurcation in correlation length of the Ising model on a ``Toblerone'' lattice}

\author{Joseph Chapman\, \orcidlink{0009-0000-7737-5837}, Bruno Tomasello\,\orcidlink{0000-0002-1156-5408}, Sam Carr \,\orcidlink{0000-0001-9995-4944}}

\address{University of Kent, Physics of Quantum Materials group\\
	Canterbury, UK, CT2 7NZ
}%

\date{\today}

\begin{abstract}
	The classical Ising chain is the paradigm for the non-existence of phase transitions in 1D systems and was solved by Ernst Ising one hundred years ago. More recently, a decorated two leg Ising ladder has received interest for the curious thermodynamics that resemble a phase transition; a sharp peak in the specific heat at low, but finite temperature. We use this model to reveal a bifurcation in the correlation lengths due to a crossing of the sub-leading eigenvalues of the transfer matrix, which results in two distinct length scales necessary to describe to the decay of correlations. We discuss this phenomenon in the context of the geometric frustration in the model. We also provide additional results to aid in the understanding of the curious thermodynamics of the model through a study of the magnetic susceptibilities.
\end{abstract}

\noindent{\it Ising Model; Phase transitions; Frustrated magnetism; Statistical Mechanics; Correlations; Low-dimensional systems \/}%
\maketitle

\section{\label{sec:intro}Introduction}

The physics of low-dimensional systems have been a constant curiosity, and vibrant area of research \cite{renard1987magnetic,moran2012physics,lipowski2022ising}. This low-dimensionality often permits exact solution of these models, allowing for probing of complex critical phenomena. Perhaps the most ubiquitous, paradigmatic example of this type is the one-dimensional Ising model, named for Ernst Ising, who first solved the model one hundred years ago \cite{1925Ising}. This model describes a chain of classical spins $s_i =\pm1$ interacting with their neighbours, with some energy scale $J$. Ising's solution of this model demonstrated that there was no phase transition at finite temperature, although his prediction that the absence of criticality would persist in higher dimensions would be disproved \cite{peierls1936ising,onsager1944crystal}.

Physically, the lack of a phase transition in one-dimension (we henceforth refer only to finite temperature transitions) can be understood by considering the energy cost, or entropy gain, for the creation of domain walls. Domain walls in a 1D system are point-like, and have a finite energy, proportional to the energy scale set by the spin-spin interaction $J$, this energy is independent of the size of the domain. However, this is in contrast to the entropy associated with these domain walls. The possible locations at which domain walls can be placed is proportional to the system size, which in the thermodynamic limit, clearly dominates the free energy. These domains can become arbitrarily large, hence preventing the formation of long range order in 1D systems. 
The 1D Ising model is exactly solved by the transfer matrix method (see textbooks on statistical mechanics~\cite{yeomans1992statistical}). By noting that the transfer matrix is positive-definite (its elements are Boltzmann weights), the lack of criticality can be understood mathematically thanks to the theorem of Perron and Frobenius for matrices \cite{perron1907theorie}. This theorem ensures that the largest eigenvalue, which controls the thermodynamics, strictly has an algebraic multiplicity of one. Level crossings of the largest eigenvalue are thus prohibited, and so the thermodynamics are smooth functions of temperature.

\begin{figure}[]
	\centering
	\includegraphics[width=0.7\textwidth]{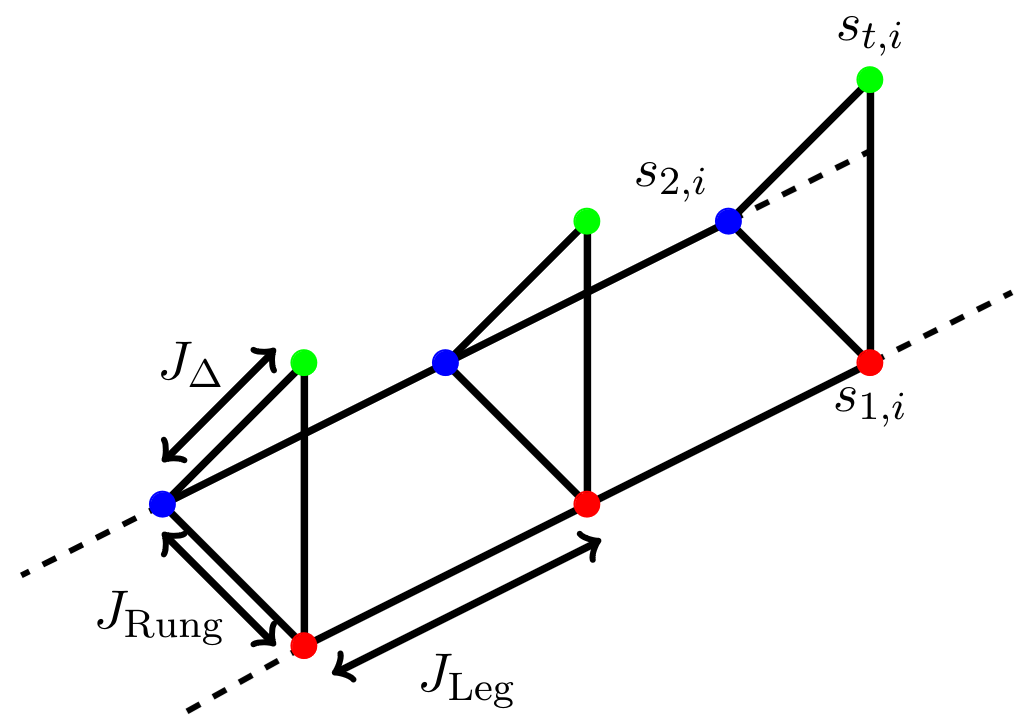}    
	\caption{ Sketch of the two leg Ising ladder with triangular rungs (or ``Toblerone" lattice), whose Hamiltonian is defined in \cref{tobl_Hamiltonian}. Spins on the legs are labelled $s_1$ and $s_2$ (red and blue dots, respectively), while $s_t$ denotes the top spins (green dots).}
\label{fig:tobl_sketch}
\end{figure}

There are a few 1D models that \textit{do} exhibit true thermodynamic phase transitions at finite temperature, see those detailed in Ref.~\cite{cuesta2004general}, however these models have transfer matrices that do not belong to the class that Perron-Frobenius applies to; usually involving long range interactions or forbidden energy states. Thus, through Perron-Frobenius, we have a strict ``no-go" theorem for the existence of phase transitions in 1D models that have positive-definite transfer matrices of finite size. 
Yet recent studies, conducted on a variety of 1D spin models, have uncovered near singular behaviour in the free energy \cite{rojas2016thermal,strevcka2016spin,de2018quasi,carvalho2018quantum,rojas2019universality,rojas2019peculiarities,carvalho2019correlation,strecka2020pseudo,krokhmalskii2021towards,sznajd2022ising}. This behaviour is not limited to spin models, and can also be observed in coupled spin-electron models \cite{galisova2015vigorous}, 1D q-state Potts models \cite{panov2021unconventional}, and spin-pseudospin models \cite{strevcka2024pseudo}. 
This manifests as thermodynamics reminiscent of a phase transition at finite temperature;  a remarkably sharp peak in the specific heat capacity, resembling a divergence as observed in cases of higher dimensionality. One specific model is a decorated two-leg Ising ladder with triangular rungs~\cite{yin2024paradigm,Yinfinding,yin2023marginal,yin2024approaching,hutak2021low}, with the Hamiltonian 
\begin{equation}\hspace{-3em}
	\mathcal{H} = -\sum_i \Big[ J_{\mathrm{Leg}}(s_{1,i} s_{1,i+1 } + s_{2,i} s_{2,i+1}) 
	+ J_{\mathrm{Rung}} s_{1,i} s_{2,i}  +  J_\Delta  (s_{1,i} + s_{2,i}) s_{t,i} \Big],
	\label{tobl_Hamiltonian}
\end{equation}
whose spins are classical ($s_{\alpha, i}=\pm 1$), and their positions and interactions are sketched in \cref{fig:tobl_sketch}. This Hamiltonian corresponds to the Ising model on the ``Toblerone" lattice,  so-called for the resemblance to the famous Swiss chocolate. $\mathcal{H}$ is invariant under the transformations $J_{\mathrm{Leg}}\to-J_{\mathrm{Leg}}$, and $J_{\Delta} \to -J_\Delta$. Interestingly, if $J_{\mathrm{Rung}} < 0$ (antiferromagnetic, AF) the spins of each $i$-th triangle become frustrated. This occurs whenever $s_{1,i}, s_{2,i}, s_{t,i}$ cannot uniquely minimise the energy of their interactions $J_{\Delta},J_{\mathrm{Rung}}$. 

Geometrical frustration -- the paradigmatic example of which is an equilateral triangle of AF interacting Ising spins -- is considered pivotal in understanding how a narrow peak can arise in the specific heat of a 1D Ising model such as $\mathcal{H}$ in \cref{tobl_Hamiltonian}~\cite{yin2024paradigm}. In general, it is well known that frustration results in manifolds of highly degenerate energy states, which in a ``perfectly" frustrated system would be the ground state. However, in the model in \cref{tobl_Hamiltonian}, this manifold occurs in the low energy states, \textit{not} the ground state . In a lattice of interconnected triangular units, the degree of degeneracy scales extensively, and the entropy associated with it can become macroscopic in the thermodynamic limit~\cite{moessner2006geometrical}. 

\begin{figure}[]
\centering
\includegraphics[width=0.7\textwidth]{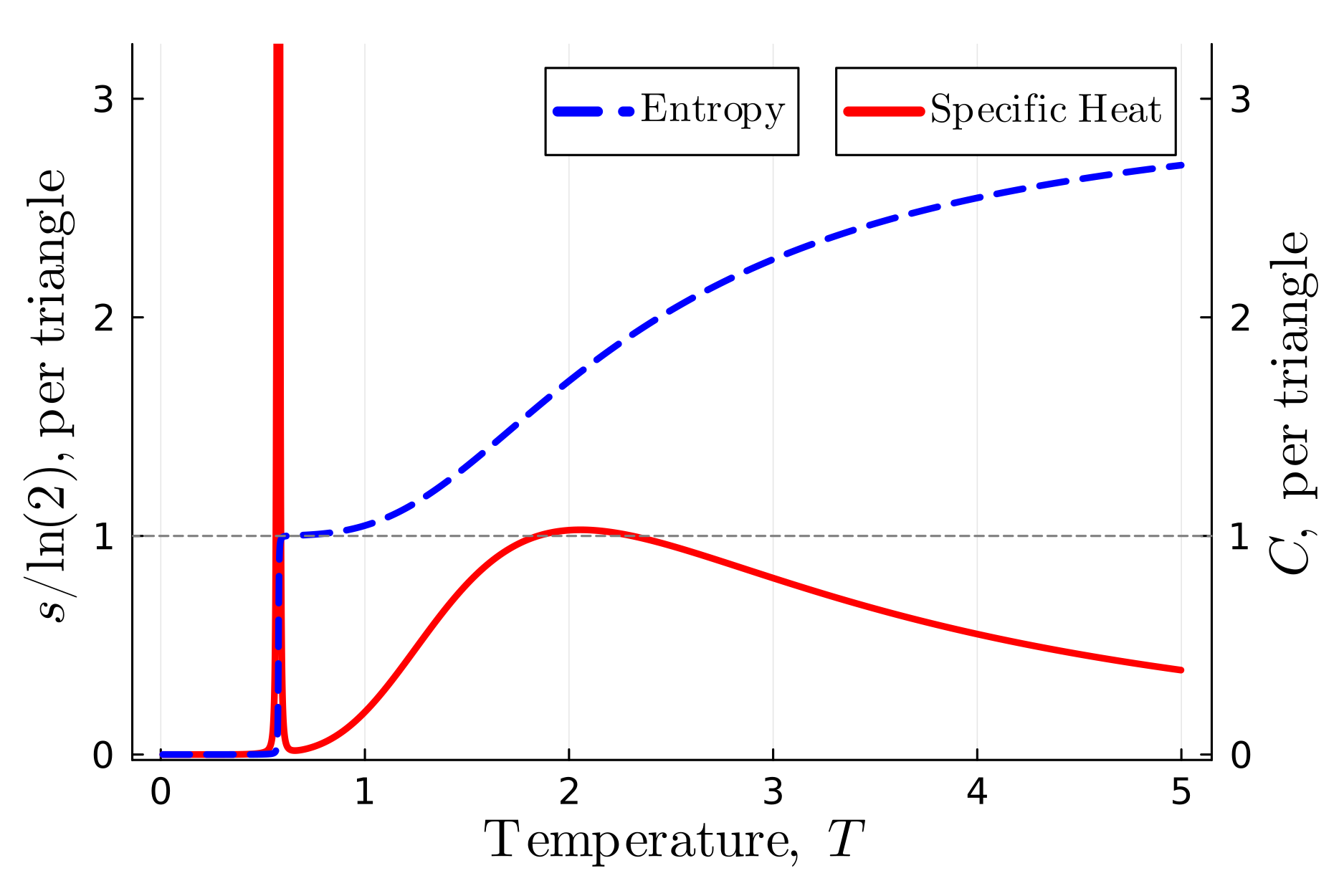}
\caption{Plot of the Specific heat and entropy for the model. The height of the peak in specific heat extends beyond the top margin, but remains of finite height and width. In this plot, $J_{\mathrm{Leg}}=2$, $J_{\mathrm{Rung}}=-1$, $J_{\Delta}=1.2$. } 
\label{fig:thermodynamics}
\end{figure}

This general connection between frustration and narrow peaks in the specific heat (often referred to as ``pseudo-transitions") was first explored in Ref.~\cite{galisova2015vigorous}, and then further generalised by a conjecture in Ref.~\cite{rojas2020conjecture}. According to this work, one must consider the ground state entropy as a function of some model parameter -- in the case of the the model defined in \cref{tobl_Hamiltonian}, this parameter is $x = J_\Delta / |J_\mathrm{Rung}|$. For $x >1$ (considered in this paper), we have an ordered ground state with zero temperature entropy $S=0$ per unit cell. However, for $x<1$ the ground state would be frustrated, and have a zero temperature entropy of $S=\ln(2)$ per unit cell. In the proximity of this boundary, the manifold of frustrated states can be accessed at low, but finite, temperature. The conjecture in \cite{rojas2020conjecture} states that this can give rise to a sharp peak in specific heat, depending on details of this zero temperature transition. These further conditions are indeed satisfied by the model in \cref{tobl_Hamiltonian}, as evidenced by the peak shown in \cref{fig:thermodynamics}.


In Ref.~\cite{yin2024paradigm}, W. Yin discussed the curious thermodynamics of the model in \cref{tobl_Hamiltonian}. It was shown that the correlation function (here denoted $\Gamma_{1,2}$) between the rung spins $s_{1}, s_{2}$ of the same triangle exhibits a \textit{change of sign} at a temperature that coincided with the temperature of the peak in the specific heat (from now on we label such temperature as $T_c$). At low temperature, the correlation function is positive, $\Gamma_{1,2} > 0$, yet upon increasing temperature, the spins along the legs of the ladder become uncorrelated ($\Gamma_{1,2}=0$), at exactly $T_c$. The correlation function changes sign upon further increasing temperature, and now assumes negative values, $\Gamma_{1,2} < 0$. This corresponds to changing from parallel to anti-parallel the orientation of the rung spins $s_{1,i}, s_{2,i}$. 
The link between a change of sign of the short range correlation functions and frustration had been discussed in Ref.~\cite{dos1992temperature}. The temperature at which the correlation function changes sign defines a "frustration temperature", which provides an immediate link to the work of Ref~\cite{yin2024paradigm}.
Indeed this sheds light on the rise of the peak in the specific heat, signalling the larger entropy of the anti-parallel, frustrated, configurations. Put simply, there are many more configurations with $s_{1,i}= -s_{2,i}$ than there are with $s_{1,i}= s_{2,i}$, and this is because $J_{\mathrm{Rung}} < 0$ competes with $J_\Delta$ frustrating their interaction with $s_{t,i}$.

This idea was elucidated upon further by Hutak \textit{et al.} by means of a renormalisation group (RG) analysis~\cite{hutak2021low}. By tracing out the top spins ($s_{t,i}$), the legs $s_{1,i} $ and $s_{2,i}$ of the ladder interact via a temperature dependent effective interaction,  $J_{\perp}(T)$, which is defined as 
\begin{equation}\label{RG_eq}
J_\perp(T) = J_{\mathrm{Rung}} + \frac{T}{2} \ln \cosh \frac{2J_\Delta}{T},
\end{equation}
and is found to change sign at the same $T_c$ estimated by Yin in Ref.~\cite{yin2024paradigm}. This change of sign conveys an effective decoupling between the legs 1 and 2 of the ladder ($J_\perp=0$), whereupon above $T_c$, the legs re-couple with an antiferromagnetic interaction ($J_\perp < 0$). The effect of the spins $s_{t,i}$ can be seen to be analogous to a variable $J_{\mathrm{Rung}}$ in a standard two-leg Ising ladder. These RG results corroborated the analysis by Yin based on the correlation functions.

Ultimately, these behaviours observed in the thermodynamics signal a \textit{crossover} from a non-degenerate ground state to a manifold of frustrated states, which, in the case of the model in \cref{tobl_Hamiltonian}, can become remarkably narrow \cite{yin2024paradigm}. These analyses provided useful insights into the curious thermodynamics, yet there remain some questions which we aim to address. 

\paragraph*{(1)} Are the location and the width of the peak in specific heat independent?
\paragraph*{(2)} What response, if any, is seen in the zero field susceptibility for this model?
\paragraph*{(3)} Can the relevant symmetries of the model provide deeper insights about the model?
\paragraph*{(4)} What is the full picture of the correlations in the model?
\\

\paragraph*{Outline of the paper --}
We address each of these questions in the following sections of the present work, which are organised as follows. We first discuss the characteristics of the peak in the specific heat, namely the location and the width, in \cref{pk_width_loc}. This section will show that the location and the width of the peak are controlled by independent parameters. Then, \Cref{susceptibility} studies the magnetic susceptibility, in zero field, for the model, and discusses the behaviour in relation to the Curie-Weiss law. We will use these susceptibilities to describe the underlying mechanisms of the crossover phenomena. \Cref{eigenvals_analytics} will study the eigenvalues of the transfer matrix more closely, considering the behaviour of the whole eigenvalue spectrum. We find that there is a crossing of the sub-leading eigenvalues, and we discuss the consequences of such, and show how this crossing is related to the symmetries of the model. The consequences are seen immediately in \Cref{correlations}, where the correlations are studied. This section presents a remarkable phenomenon, in which we find a bifurcation in the correlation lengths, in part, as the result of a crossing of eigenvalues of the transfer matrix. This bifurcation is discussed in the context of the frustration of the spins $s_{t,i}$, and we show that this bifurcation persists even under the inclusions of an interaction between the spins $s_{t,i}$. Including this additional interaction aids in probing the link between the thermodynamics and the correlation length bifurcation.

\section{\label{pk_width_loc} Control of the peak in the specific-heat}

In order to study the specific heat, it is necessary to first obtain the free energy. This can be obtained by the transfer matrix method \cite{yeomans1992statistical}, through which we obtain the free energy $F/N = -T \ln \lambda_1$, ($\lambda_1$ is the dominant eigenvalue). The transfer matrix, and additional details can be found in \ref{app_TM}. From the free energy, we can compute the entropy by $s = -\partial f / \partial T$, and then the specific heat from $c = T \partial s / \partial T$. 

The ability to control the characteristic features of the peak is pivotal in assessing the physical foundations of the model. We parametrise the peak in the specific heat by the temperature at which the peak is centred ($T_c$) and by the width of the peak ($\Delta T$). (While this manuscript was in preparation, W. Yin addressed the width of the peak \cite{yin2024paradigm}. The result we present is in good agreement with that found by W. Yin). 
A most relevant point is to establish whether there are sets of parameters from \cref{tobl_Hamiltonian} which control  $T_c$ and $\Delta T$ independently, since the narrow width of the peak could otherwise be related to the temperature where the peak occurs. 

\begin{figure}[h!]
	\begin{subfigure}[c]{0.5\linewidth}
		\centering
		\includegraphics[width=\linewidth]{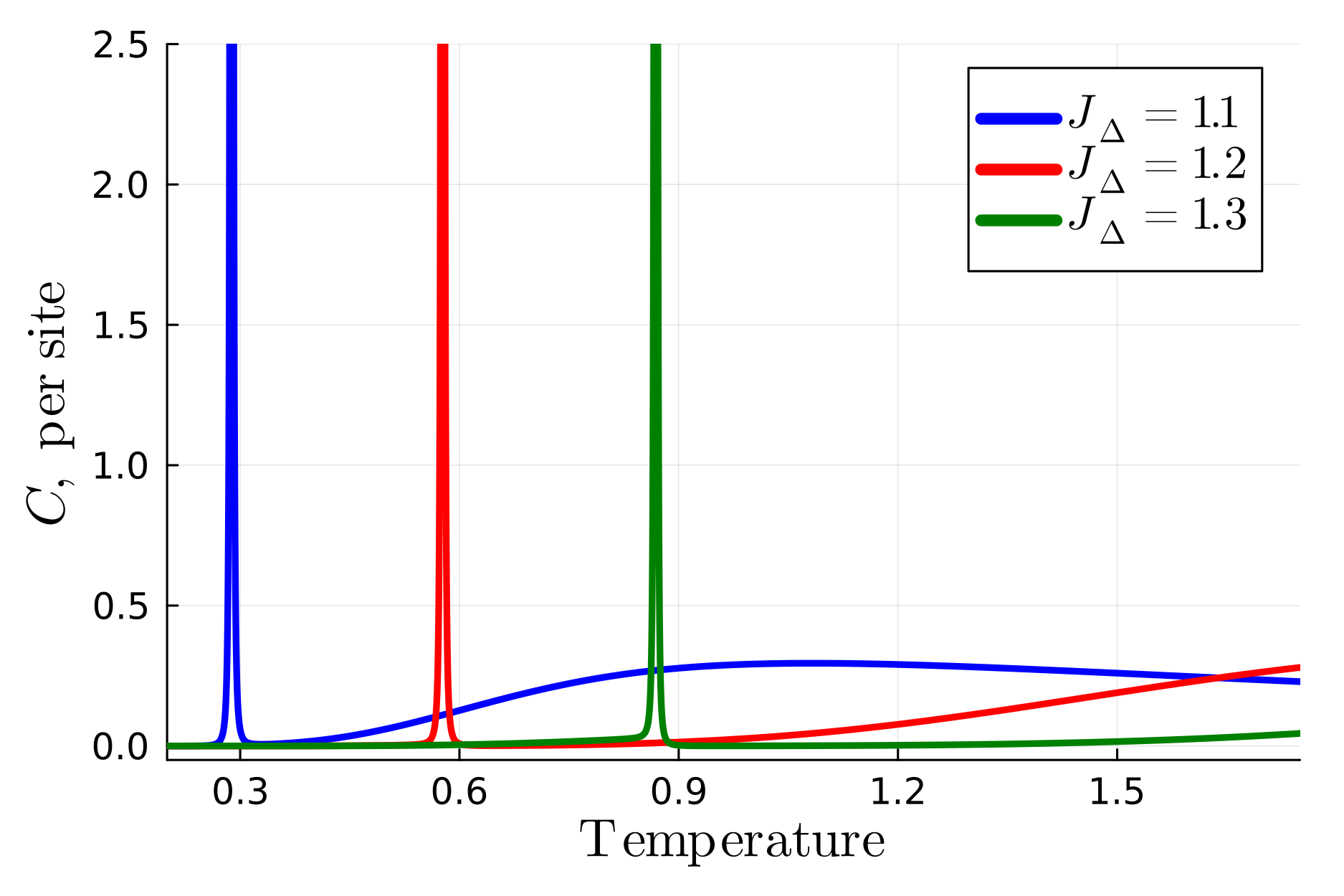}
		\caption{}
		\label{fig:pk_loc}
	\end{subfigure}
	\begin{subfigure}[c]{0.5\linewidth}
		\centering
		\includegraphics[width=\linewidth]{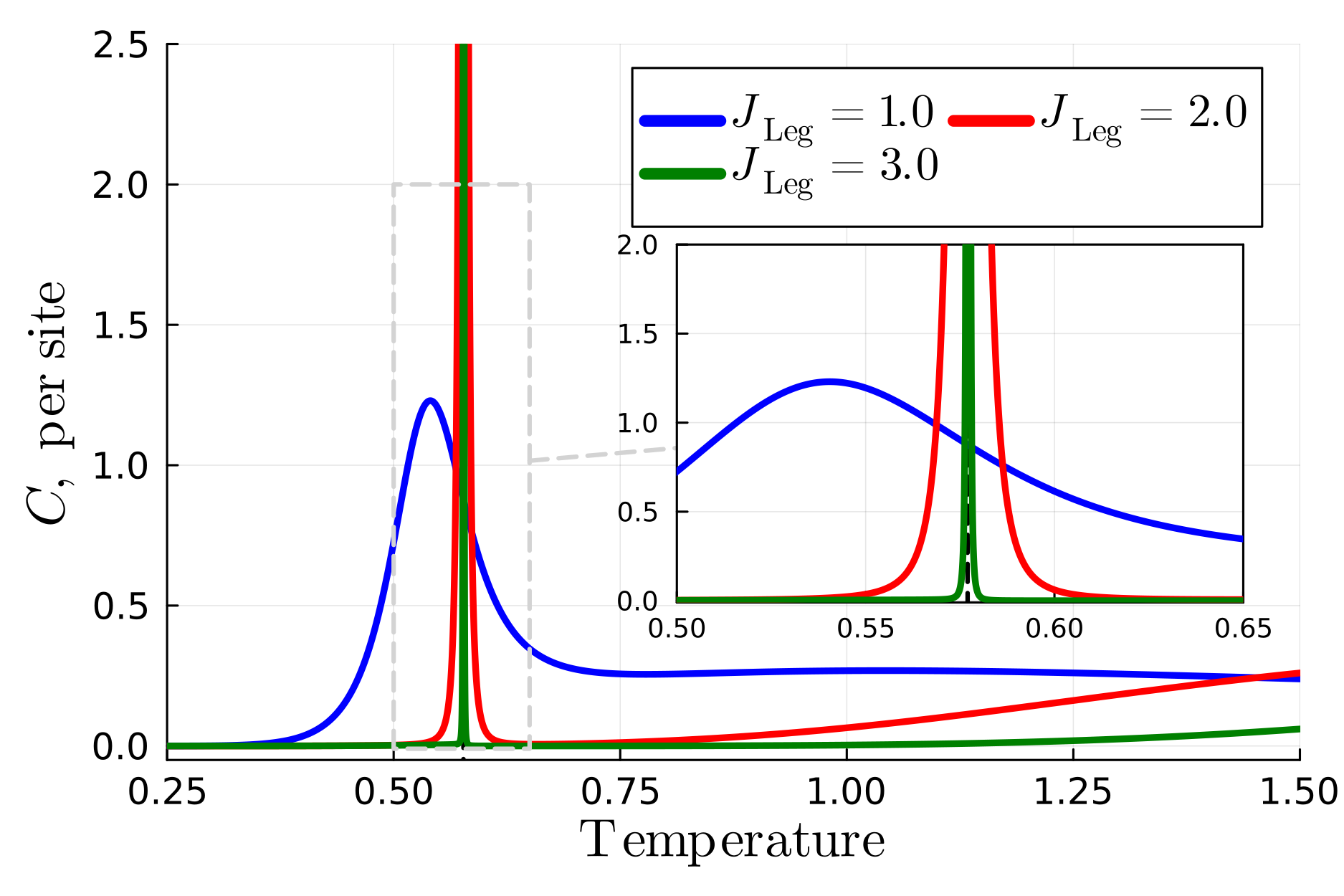}
		\caption{}
		\label{fig:pk_width}
	\end{subfigure}
	\label{fig:w_loc}
	\caption{Control over the location [panel (a)] and the width [panel (b)] of the peak in the specific heat through different parameters. (The height of the sharp peaks extends past the top of the plot.) Panel (a): the peak location is controlled by changing $J_\Delta$, and keeping $J_{\mathrm{Rung}}=-1$. Panel (b): the peak width is parametric in $J_\mathrm{Leg}$, while $J_\mathrm{Rung}=-1, J_\Delta=1.2$ are fixed, yielding $T_c \approx 0.577$ (vertical dashed line, also in the inset). We note that for $J_\mathrm{Leg}=1$ in (b), the peak is much broader, and no longer centred at $T_c$.}
\end{figure}

\paragraph*{Peak position --}
The position of the peak has been well documented in current literature ~\cite{yin2024paradigm,hutak2021low}. In Ref.~\cite{yin2024paradigm}, the position of the peak is obtained as the temperature at which a so-called ``frustration function" $I_-$ is zero (the exact form of $I_-$ is given in \cref{app_TM}). In a similar fashion, Ref.~\cite{hutak2021low} obtains the same position of the peak by considering the temperature at which the effective coupling $J_\perp(T)$, as given in \cref{RG_eq}, is zero. We reproduce the result, which is
\begin{equation}\label{tc_eq_lit}
T_c \approx \frac{2}{\ln 2}(|J_\Delta| + J_{\mathrm{Rung}}),
\end{equation}
which gives $T_c$ as dependent only on the parameters $J_{\Delta}, J_{\mathrm{Rung}}$ and independent of $J_\mathrm{Leg}$. To illustrate this result, the specific heat is plotted for various values of $|J_\Delta| + J_{\mathrm{Rung}}$ in \cref{fig:pk_loc}. For the three values of $|J_\Delta|= 1.1, 1.2, 1.3$, used in \cref{fig:pk_loc}, the corresponding calculated values of $T_c \approx 0.2885,0.5771,$ and $0.8656$ are found. Each of these peaks has a width much narrower than the corresponding peak position ($\Delta T\ll T_c$). We thus clearly see that by tuning $J_{\Delta}, J_{\mathrm{Rung}}$, we are able to arbitrarily place the peak. 
What has been less well documented, however, is the width of this peak. It is necessary to discern the parameters that control the width of this peak, as mentioned previously. 

\paragraph*{Peak width --}
The peak width can be studied by expanding the specific heat around $T_c$. The dominant contribution to the specific heat, in the vicinity of $T_c$, is given conveniently in Ref.~\cite{hutak2021low} in terms of the renormalised interaction $J_\perp(T)$. This dominant contribution is
\begin{equation}\label{c_rg_expr}
C \sim  const.\ \frac{\cosh(J_\perp / T) \cosh(2J_\mathrm{Leg} /T)}{(1 + \sinh^2(J_\perp / T) \cosh^2(2J_\mathrm{Leg} /T))^{3/2}},
\end{equation}
where the constant is related to $\partial J_\perp(T) / \partial T$. By then considering the behaviour near $T_c$, this expression can be expanded as
\begin{equation}
C \sim \frac{1}{[1 + (T-T_c)^2/\eta]^{3/2}},
\end{equation}
where $\eta = [ 2\: T_c  \cosh(2J_{\mathrm{Leg}}/T_c) /\ln(2) ]^2$. We consider the point at which the specific heat reaches half its maximum value, the full width at half maximum. We write this point as
\begin{equation}\label{eq_fwhm}
\frac{C}{C_{\mathrm{max}}} = \frac{1}{2}=\frac{1}{(1+A)^{3/2} },
\end{equation}
where $A = (\Delta T^2 / 2 )\ \eta^{-1}$. We find that $A = 2^{2/3} - 1$ from \cref{eq_fwhm}, and by equating $(\Delta T^2 / 2 )\ \eta^{-1} =2^{2/3} - 1$ we can obtain an expression for the width which is
\begin{equation}\label{width_eq}
\Delta T= \alpha_{\Delta T} \, \frac{T_c}{\ln 2}\ e^{-2J_{\mathrm{Leg}}/T_c},
\end{equation}
where $\alpha_{\Delta T} = 4\sqrt{2 (2^{2/3} -1)}$, and we make use of $e^{J_{\mathrm{Leg}}/T_c} \gg e^{-J_{\mathrm{Leg}}/T_c}$. For a fixed value of $T_c$, the width $\Delta T$ decreases exponentially with increasing $J_{\mathrm{Leg}}$. This is illustrated in \cref{fig:pk_width} for three values of $J_{\mathrm{Leg}}$, and a constant $T_c \approx 0.577$. 

Thus the location and width of the peak can be controlled by independent parameters of the model, i.e., $T_c$ is governed solely by $J_\Delta$ and $J_{\mathrm{Rung}}$ while $\Delta T$ by $J_\mathrm{Leg}$ for a fixed value of $T_c$. It is relevant to ask whether $\Delta T$ can become zero, as this would correspond to a phase transition. However, \cref{width_eq} is zero only when $T_c = 0$, corresponding to the phase transition at zero temperature in the 1D Ising Model. Whilst this peak, and hence the crossover, can become remarkably narrow at finite temperature ($\Delta T \ll T_c$), it is of strictly finite height and width.  

For completeness, we include the case  $J_{\mathrm{Leg}}=1$ in \cref{fig:pk_width},  where the width of the peak is now of the order $\Delta T = \mathcal{O}( |J_\Delta| + J_{\mathrm{Rung}})$. The peak location now deviates from the calculated $T_c$, and is much broader. We will return to this broadening of the peak in\cref{correlations}.  

\section{\label{susceptibility} Zero Field Susceptibility}
This section considers the zero-field magnetic susceptibility of the model defined in \cref{tobl_Hamiltonian}. We will study four key susceptibilities, and discuss the relation to the crossover phenomena in the model. These are a uniform field to each spin, $\chi_{\mathrm{Uniform}}$; next, we apply a field uniformly to only the spins on the legs, which we refer to as the ``ferromagnetic" susceptibility , $\chi_{\mathrm{FM}}$; then applying the field in opposite directions to the spins on the two legs of the ladder, the ``antiferromagnetic" susceptibility, $\chi_{\mathrm{AFM}}$; and finally a field applied to only the top spins, $\chi_{\mathrm{Top}}$.

To study these four cases, we include the corresponding field term in the Hamiltonian for the model, \cref{tobl_Hamiltonian}, which are
\begin{equation}\label{chi_eqs}
	\eqalign{\chi_{\mathrm{Uniform}} &\to \quad \mathcal{H}_{B} = -B \sum_i (s_{1,i}+ s_{2,i} + s_{t,i}), \cr
	\chi_{\mathrm{AFM}} &\to \quad \mathcal{H}_{B} = -B \sum_i (s_{1,i}- s_{2,i}), \cr
	\chi_{\mathrm{FM}} &\to \quad \mathcal{H}_{B} = -B \sum_i (s_{1,i}+ s_{2,i}),\cr
	 \chi_{\mathrm{Top}} &\to \quad \mathcal{H}_{B} = -B \sum_i s_{t,i}.	}
\end{equation}

After making the necessary modifications, the transfer matrix can no longer be block diagonalized as the symmetry under a total spin-flip is no longer present. As such, we make use of a centred finite difference method in order to compute the zero-field susceptibilities numerically.

\begin{figure}
	\centering
	\includegraphics[width=0.75\linewidth]{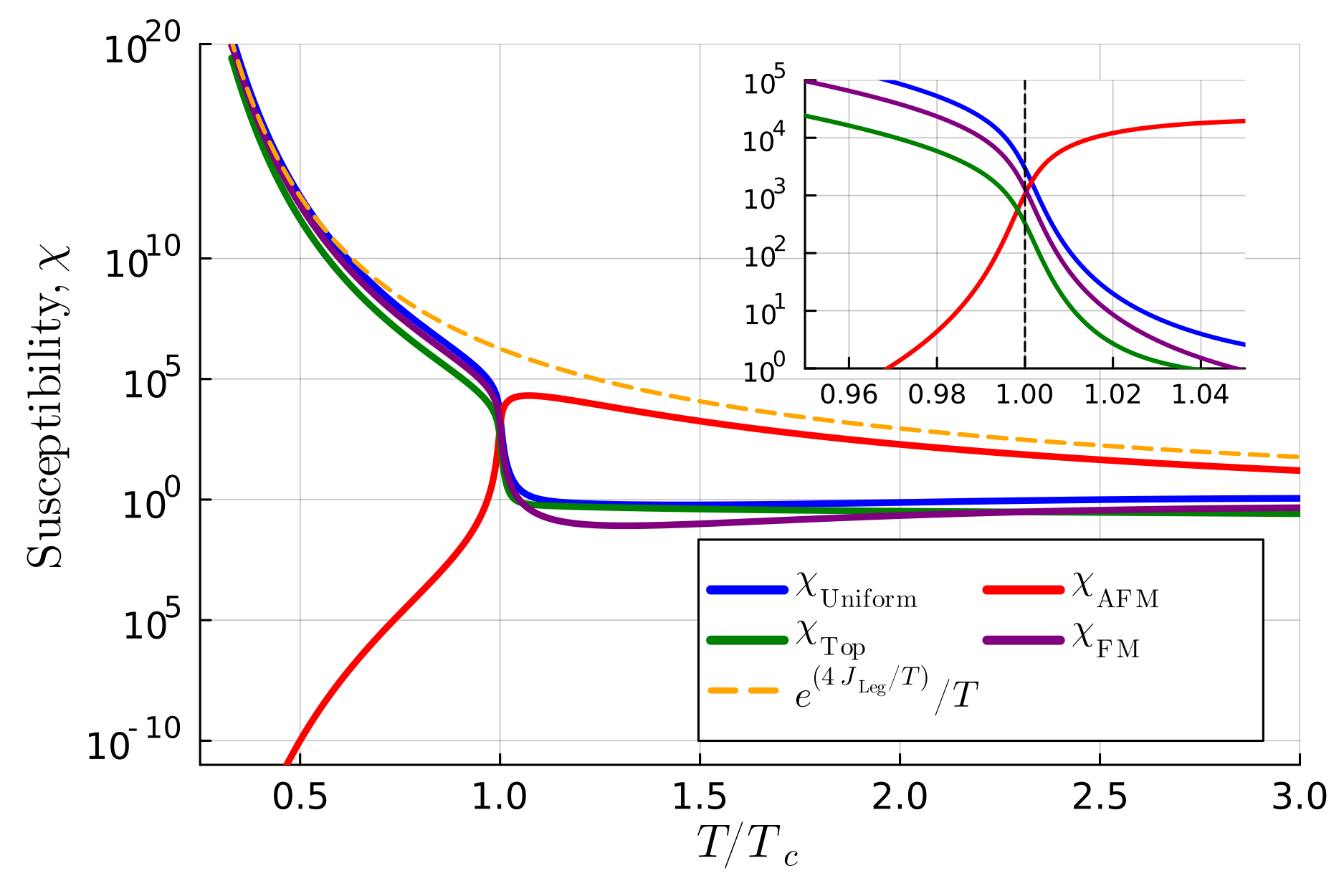}
	\caption{Plot of the susceptibilities $\chi$, as a function of temperature. For these plots, we have used the parameters $J_{\mathrm{Leg}}=2, J_{\mathrm{Rung}}=-1,J_\Delta=1.2$. The dashed line shows $\exp(4J_{\mathrm{Leg}}/T)/T$ as guide for the low-T behaviour. The susceptibilities of largest magnitude provide insights about the spin configurations at a given temperature. For $T<T_c$, we see that the ``ferromagnetic" susceptibilities dominate, whilst for $T>T_c$, the ``antiferromagnetic" susceptibility dominates. The inset shows a zoomed view around $T_c$. }	\label{fig:chi_plots}
\end{figure}

The four susceptibilities are plotted in \cref{fig:chi_plots} on a $\log_{10}$ scale, with an inset showing a zoomed view around $T_c$. Two distinct behaviours of the susceptibilities can be observed in \cref{fig:chi_plots}; for $T < T_c$, $\chi_{\mathrm{Uniform}},\chi_{\mathrm{FM}},\chi_{\mathrm{Top}}$ all become exponentially large. This exponential behaviour is shown by the dashed line in \cref{fig:chi_plots}, which corresponds to $\exp(4\ J_{\mathrm{Leg} }/T) /T$, which is an estimate of two FM Ising chains. For reference, the standard Ising chain has a low temperature susceptibility that behaves as $\chi \sim \exp(2J / T) /T$. The magnitudes of these susceptibilities for $T < T_c$ indicates that parallel configurations of spins is preferred; we observe $\chi_{\mathrm{AFM}}$ becoming exponentially small below $T_c$ which indicates that anti-parallel alignments are suppressed. However, this changes for $T>T_c$, where we observe that $\chi_{\mathrm{AFM}}$ is now the dominant susceptibility, which conveys a preference for configurations with $s_{1,i} = -s_{2,i}$. 

Together, these susceptibilities corroborate the understanding of the crossover mechanism.  At low temperature, $T<T_c$, each triangular unit has its spins aligned parallel, this is at the expense of satisfying $J_{\mathrm{Rung}}$, however this is energetically favourable owing to $|J_\Delta| > |J_{\mathrm{Rung}}|$. Upon increasing temperature, this preference toward parallel alignments becomes suppressed, and instead yielding to anti-parallel alignments. The effect of this, is that the orientation of $s_{t,i}$ is no longer parallel to both $s_{1,i},s_{2,i}$; it has become \textit{frustrated}, resulting in an entropy of $\ln2$ per triangle, which gives rise to the peak in the specific heat. 

\begin{figure}
	\centering
	\includegraphics[width=0.75\linewidth]{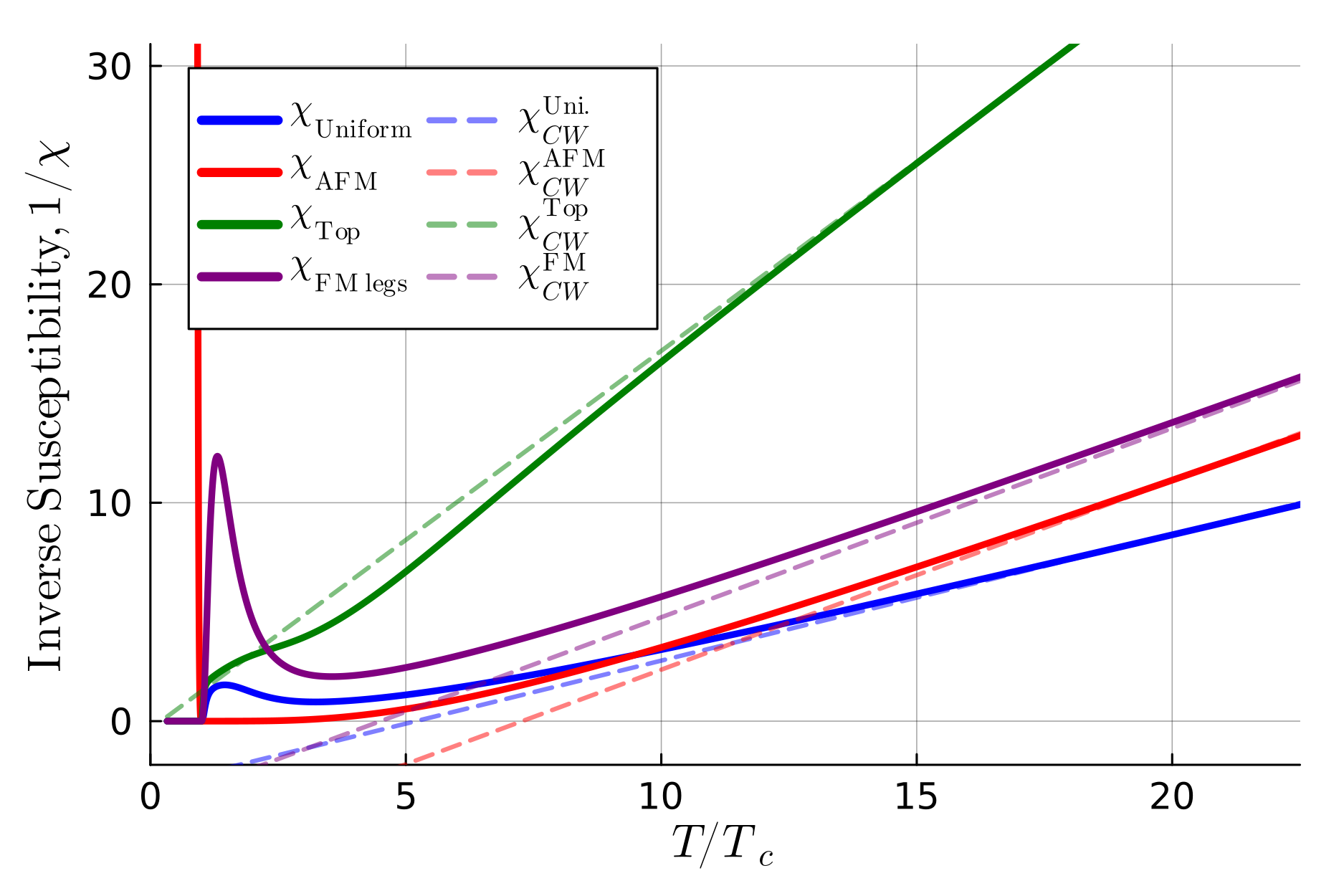}
	\caption{Plot of the inverse susceptibilities $1 / \chi$, as a function of temperature. For these plots, we have used the parameters $J_{\mathrm{Leg}}=2, J_{\mathrm{Rung}}=-1,J_\Delta=1.2$. Each solid line ($1/\chi$) has a corresponding dashed line that shows the Curie-Weiss result for high temperature, with which we see good agreement. The specifics of the mean field analysis is discussed in the main text.}
	\label{inv_chi_plots}
\end{figure}

We also perform a mean field analysis in order to make a comparison between the Curie-Weiss law and our numerical results. Self-consistent equations for the on site magnetisations are formed for the four cases given in \cref{chi_eqs}, which are then solved to obtain expressions for the magnetic susceptibility, assuming a linear response to the field. 
Generally, the inverse susceptibility behaves linearly at high temperatures, and this region can be well approximated by the equation
\begin{equation}\label{chi_cw}
	\chi_{CW} \sim  n\ \frac{1}{T- \theta_{CW}},
\end{equation}
where $\theta_{CW}$ is the Curie-Weiss temperature, and $n$ is a normalisation constant. For each of the cases in \cref{chi_eqs}, values of $\theta_{CW}$ and $n$ obtained from the mean field analysis are tabulated in \cref{CW_results}. We see good agreement between the results in \cref{CW_results} and the numerical results in \cref{inv_chi_plots} at high temperature. 

\begin{table}
\caption{\label{CW_results}Tabulated results for the mean-field analysis of the inverse susceptibilities in \cref{chi_cw}.}
\footnotesize\rm
\begin{tabular*}{\textwidth}{@{}l*{15}{@{\extracolsep{0pt plus10pt}}l}}
	\br
	Susceptibility& n & $\theta_{CW}$ \\
	\mr
	$\chi_{\mathrm{Uniform}}$ & 1 & $2J_{\mathrm{Leg}} + J_{\mathrm{Rung}}$\\
	$\chi_{\mathrm{FM}}$ & 2/3 & $ 2J_{\mathrm{Leg}} + J_{\mathrm{Rung}}- J_\Delta$\\
	$\chi_{\mathrm{AFM}}$ & 2/3 & $2J_{\mathrm{Leg}} - J_{\mathrm{Rung}}+2J_\Delta$\\
	$\chi_{\mathrm{Top}}$ & 1/3 & $2J_{\mathrm{Leg}} + J_{\mathrm{Rung}}-2J_\Delta^2$\\
	
\end{tabular*}
\end{table}

\section{\label{eigenvals_analytics}Transfer Matrix: Symmetries \& level crossings}
Let us now consider the full structure of the transfer matrix in detail.  As discussed, the thermodynamics depend only on the largest eigenvalue of the transfer matrix $\lambda_1$, in the thermodynamic limit. The correlations, however, depend on the sub-leading eigenvalues, and so it is prudent to study the entire eigen-spectrum more closely.
For the model in \cref{tobl_Hamiltonian}, the transfer matrix has dimensions of $8\times8$ (three Ising spins per unit cell), and its symmetries permit block diagonalisation, which allows analytical expressions for the eigenvalues to be obtained~\cite{yin2024paradigm,mejdani1996vII}. We find that four of the eigenvalues are exactly zero, directly relating to tracing out the spins $s_{t,i}$, and hence obtaining a $4\times4$ matrix. The non-zero eigenvalues of the transfer matrix ($\lambda_i$) are given in \ref{app_TM}, along with further details related to the rest of this section.

\begin{figure}[htbp]
\centering
\includegraphics[width=0.75\textwidth]{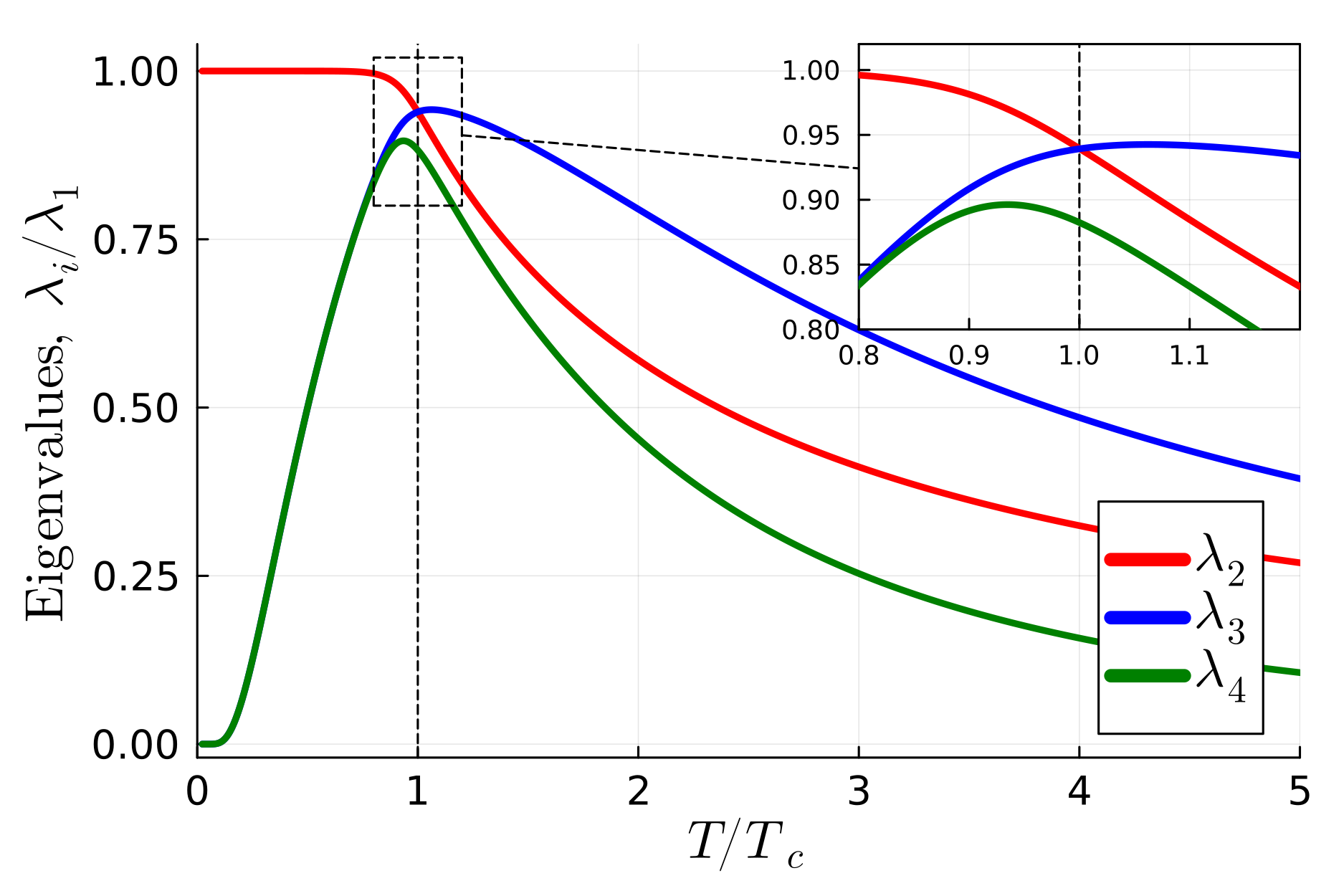}
\caption{The behaviour of the four non-zero, sub-leading eigenvalues, $\lambda_{2,3,4}$, divided by $\lambda_1$. The vertical line is a guide for the eyes to highlight the temperature where $\lambda_2,\lambda_3$ cross. The parameters used are $J_{\mathrm{Leg}}=1,J_{\mathrm{Rung}}=-1,J_\Delta =1.2$. }
\label{fig:eigenvalues}
\end{figure}

The Perron-Frobenius theorem ensures that $\lambda_1$ never crosses any $\lambda_{i>1}$, thus leading to thermodynamic functions continuous in temperature, i.e., a ``no-go''  for thermodynamic phase transitions~\cite{perron1907theorie}. In the rest of this section, we will discuss the importance of the \textit{sub-leading} eigenvalues, specifically $\lambda_2,\lambda_3$, to the physics of the model. We find that the symmetries of the transfer matrix permit these eigenvalues to \textit{cross} as a function of temperature.  For clarity, our convention assigns labels to the $\lambda_i$ based upon the low temperature magnitudes (see \ref{app_TM}; for example, $\lambda_2 > \lambda_3$ at temperatures below the crossing point, but $\lambda_2 < \lambda_3$ at temperatures above it). This crossing, shown in \cref{fig:eigenvalues}, has profound consequences for the correlations, which are discussed in the following section.

\paragraph*{Crossing of the sub-leading eigenvalues --}

The symmetries of the transfer matrix permit a block diagonal form to be obtained, which was noted in Ref~\cite{yin2024paradigm}, although the nature of these symmetries was not fully explored. As such, we study these symmetries more closely, and demonstrate how they relate to the crossing of $\lambda_{2}$ and $\lambda_3$.

The first symmetry is the ``spin-flip" symmetry; the transfer matrix is invariant under the transformation $\ket{+ + +} \to \ket{- - -}$, with the spins entries corresponding to the vector of spin variables $\ket{s_{1}, s_{2}, s_{t}}$, which is not a quantum mechanical ket. The corresponding operator, $\hat{P}_1$, is \textit{anti-diagonal}, and has eigenvalues $\pm1$, each with multiplicity four. The matrix for this operator, and more comprehensive details can be found in \ref{app_sym}. We can see the eigenvalues clearly by considering the action of this operator on linear combinations of basis states, namely
\begin{equation}
	\eqalign{	+1: \quad \hat{P}_1 & (\ket{+ + +} + \ket{- - -}) = \ket{+ + +} + \ket{- - -}, \cr 
		-1: \quad \hat{P}_1 & (\ket{+ + +} - \ket{- - -}) = -(\ket{+ + +} - \ket{- - -}),}
\end{equation}
This operator is indeed a symmetry of the transfer matrix, as can be shown by the commutator $[\hat{P}_1,\mathbf{T}]=0$. This has the consequence that a simultaneous diagonalisation could be performed with some unitary matrix $U$. We note that such a matrix would bring the transfer matrix to a block diagonal form, with a two-block structure (each of dimension $4\times4$).  These blocks correspond to the $\pm$ parity sectors of the operator $\hat{P}_1$. This symmetry alone does not shed much light on the crossing of the sub-leading eigenvalues, as $\lambda_2, \lambda_3$ are located in the same block (corresponding to  $\hat{P}_1=-1$). To permit the crossing of $\lambda_2, \lambda_3$, they must be located in separate blocks, motivating the introduction of a second symmetry.

The second symmetry corresponds to the exchange of the legs of the ladder. The corresponding operator, $\hat{P}_2$, acts trivially on states with $s_{1,i}=s_{2,i}$, otherwise exchanging the legs, for example $\hat{P_2}\ket{+ - -} = \ket{- + -}$. This operator, like $\hat{P}_1$, has eigenvalues of $\pm1$, however, the multiplicities are now six, for the eigenvalue $+1$, and two, for the eigenvalue $-1$. Again, we can see these eigenvalues through the action of $\hat{P}_2$ on linear combinations of basis states. 

\begin{equation}
	\eqalign{+1: \quad \hat{P}_2  \left(\ket{+ - +} + \ket{- + -} + \ket{+ - -} + \ket{- + +}\right) \cr
		\qquad =  \ket{- + +} + \ket{+ -  -} + \ket{- + -} + \ket{+ - +},\cr
		-1: \hat{P}_2   \left(\ket{+ - +} - \ket{- + -} + \ket{+ -  -} - \ket{- + +} \right)\cr 
		\qquad=  - \left(\ket{+ - +} - \ket{- + -} + \ket{+ -  -} - \ket{- + +}\right) .}
\end{equation}
We can also find that $[\hat{P_2},\mathbf{T}]=0$, confirming that the transfer matrix is symmetric under the exchange of the legs. 

It is important to see that $[\hat{P_1},\hat{P_2}]=0$, meaning we are able to diagonalise $\hat{P}_1,\hat{P_2}, \mathbf{T}$ simultaneously with a single unitary matrix $U$, which can be constructed by the shared eigenstates of the three matrices. The exact form of the matrix $U$ can be found in \ref{app_sym}, along with the result of the product $U^{-1}\ \mathbf{T}\ U$, which is again a block diagonal matrix. Note that the unitary matrix used here differs from the one in \cite{yin2024paradigm}. We find that our matrix provides clearer insights for the study of the sub-leading eigenvalue crossing. Indeed, the structure of these blocks is much more informative. Four blocks are now present; two $3\times 3$ blocks, and two $1 \times 1$ blocks. These arise from the $4\times 4$ blocks, due to $\hat{P_1}$, being split by the second symmetry, $\hat{P_2}$.  This structure makes it straightforward to obtain the eigenvalues, which reveals that the eigenvalues $\lambda_2,\lambda_3$ are now located in separate blocks; $(\hat{P_1},\hat{P_2}) = (-1,+1)$, and $(\hat{P_1},\hat{P_2}) = (-1,-1)$ respectively. 

These eigenvalues cross as a function of temperature, and so we can consider the temperature at which this crossing occurs. By equating the analytical expressions of the eigenvalues, we obtain a transcendental equation that provides condition for the crossing of $\lambda_2,\lambda_3$ (see \ref{app_TM}). This expression is
\begin{equation}\label{tc_cond_cross}
	{\frac{e^{2 J_{\mathrm{Rung}}/T_c}}{2}(e^{2J_{\Delta}/T_c}+e^{-2J_{\Delta}/T_c})}=1,
\end{equation}
where $T_c$ denotes the crossing temperature. Then, under the approximation $e^{2J_\Delta /T_c} \gg e^{-2J_\Delta / T_c}$, which is satisfied for $|J_\Delta| \gg T_c$, we are able to obtain:
\begin{equation}\label{tc_eq}
	T_c \approx \frac{2}{\ln 2}(|J_\Delta| + J_{\mathrm{Rung}}),
\end{equation}
which is the same obtained in Refs~\cite{yin2024paradigm,hutak2021low}. The importance of the behaviour of the sub-leading eigenvalues for locating $T_c$ has also been discussed in Refs.~\cite{pimenta2022anomalous,rojas2024unusual} in the context of different models.

We can probe this coincidence of temperature by considering the standard two-leg ladder (without the decorating spins $s_{t}$) \cite{mejdani1996ladder,mejdani1996vII}. Comparing the operators to those relevant for the conventional two-leg ladder, we find the same symmetries commute with the transfer matrix. After performing the block diagonalisation for the two-leg ladder, we obtain two $1\times1$ blocks, and a single $2\times2$ block, which corresponds to the combinations of eigenvalues of $\hat{P_1},\hat{P_2}$. We again find that the eigenvalues $\lambda_2,\lambda_3$ are located in separate blocks. 
These eigenvalues are 
\numparts
\begin{eqnarray}	\label{eq:2_leg_eig}
	\lambda_2 &=  2\ e^{J_{\mathrm{Rung}}/T} \sinh \frac{2J_{\mathrm{Leg}}}{T} , 							\label{lambda2_ladder_eq} 	\\
	\lambda_3 &= 2\ e^{-J_{\mathrm{Rung}}/T} \sinh\frac{2J_{\mathrm{Leg}}}{T},											
	\label{lambda3_ladder_eq} 	
\end{eqnarray}
\endnumparts
which become degenerate if $J_{\mathrm{Rung}}$ is taken to be zero (i.e. the case of two decoupled chains).  This is then immediately relevant for the crossing of $\lambda_2,\lambda_3$  for the Ising model on the Toblerone lattice. It was shown by Hutak et al. in Ref.~\cite{hutak2021low} that the spins $s_t$ can generate an effective, temperature dependant interaction between the legs. This interaction, $J_\perp(T)$, goes to zero at the \textit{same} temperature at which we find the eigenvalues crossing, effectively decoupling the legs. Thus we can find the parallel between the eigenvalues of the standard two-leg ladder and the eigenvalues of the ``Toblerone" model in \cref{tobl_Hamiltonian}. The effect of $s_{t,i}$ is an effective decoupling of the legs, which allows the eigenvalues $\lambda_{2},\lambda_3$ to cross as a function of temperature. The importance of this crossing becomes apparent when studying the correlations which are known to depend on the sub-leading eigenvalues of the transfer matrix.

\section{\label{correlations} Correlations}
We now look at the spin-spin correlation functions for the model, which allows a physical understanding of the model through the detailed site-by-site information. In \cref{sec:intro,susceptibility} of this work, the rapid change in the entropy, and hence the peak in the specific heat, originates from the spins $s_{t,i}$ becoming frustrated, yielding an entropy of $s = \ln 2$ per unit cell. Despite this, Refs.~\cite{yin2024paradigm,hutak2021low} had not considered the correlations between these spins, providing immediate motivation to study the correlations more closely. Motivation also arises from the crossing of the sub-leading eigenvalues presented in the previous section. The correlations functions are known to depend on the second largest eigenvalue, and so we study the effect this crossing has.

To begin, the correlation between two spins is defined as:

\begin{equation}\label{therm_avg_corr}
	\Gamma(R) = \braket{s_0 s_R} - \braket{s_0}\braket{s_R},
\end{equation}
where $\braket{...}$ denotes a thermal average, and $R$ is the separation between the spins $s_0, s_R$. For the consideration of 1D models, $\braket{s_0}\braket{s_R} = 0$, owing to the lack of spontaneous magnetisation. 
We extend the definition in \cref{therm_avg_corr}, writing the spin-spin correlation function as
\begin{equation}
	\Gamma_{\alpha,\beta}\ (R) = \sum_{i>1} \left( \frac{\lambda_i}{\lambda_1} \right)^R \braket{u_1|\mathbf{S}_{\alpha}|u_i}\braket{u_i|\mathbf{S}_\beta|u_1} =  \sum_{i>1} e^{-R/\xi_i}  C_{\alpha,\beta}^{(i)} ,\label{corr_defn}
\end{equation}
where the largest $\xi_i$ is the correlation length, as this term describes the slowest decay of correlations. We will examine this in more detail in\cref{corr_len}. The terms $C_{\alpha,\beta}^{(i)}$ are the prefactors corresponding to the product of matrix elements, and feature the eigenvectors $\ket{u_i}$ of the transfer matrix. Additional subscripts, $\alpha,\beta$, are introduced to distinguish between the three sites in the unit cell; $\alpha,\beta= 1,2,t$ which correspond to the two legs of the ladder, and the ``top'' spins respectively.  
The $R-$dependence of \cref{corr_defn} is wholly contained in the exponential term, and the site information is contained in the prefactor term. The matrices, $\mathbf{S}_{\alpha,\beta}$ (shown in \ref{app_spin}), are diagonal, and can be deduced from the basis states of the transfer matrix. We now consider each of the terms in \cref{corr_defn} in turn.

\subsection{The prefactors} 

The prefactors, $C_{\alpha,\beta}^{(i)}$, take values in the range $C_{\alpha,\beta}^{(i)}\in [-1,1]$, and are independent of $R$. These prefactors convey the relative alignment of the spins $s_{\alpha}$ and $s_\beta$;  $C_{\alpha,\beta} > 0$ corresponds to a \textit{parallel} alignment, and $C_{\alpha,\beta} < 0$ corresponds to \textit{anti-parallel} alignments. We will focus on $C_{\alpha,\beta}^{(2)},C_{\alpha,\beta}^{(3)}$ in this section ($\lambda_2,\lambda_3 > \lambda_4$, and hence make the only significant contributions to the correlations for $R\neq0$.).

\begin{figure*}[!htb]
	\centering
	\begin{subfigure}[b]{0.49\textwidth}
		\centering
		\includegraphics[width=\linewidth]{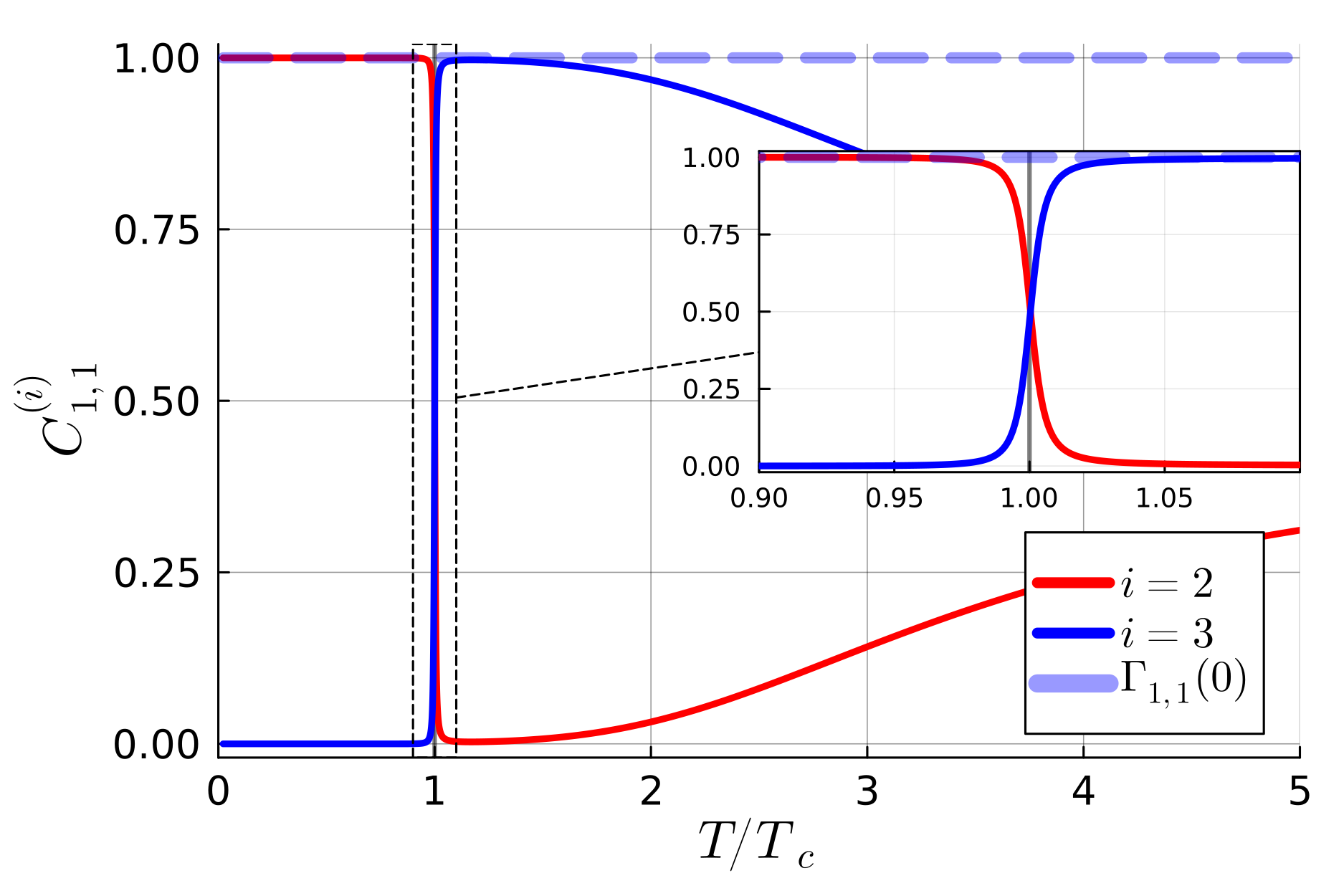}
		\caption{$C_{1,1}$}
		\label{fig:c11}
	\end{subfigure}
	\hfill
	\begin{subfigure}[b]{0.49\textwidth}
		\centering
		\includegraphics[width=\linewidth]{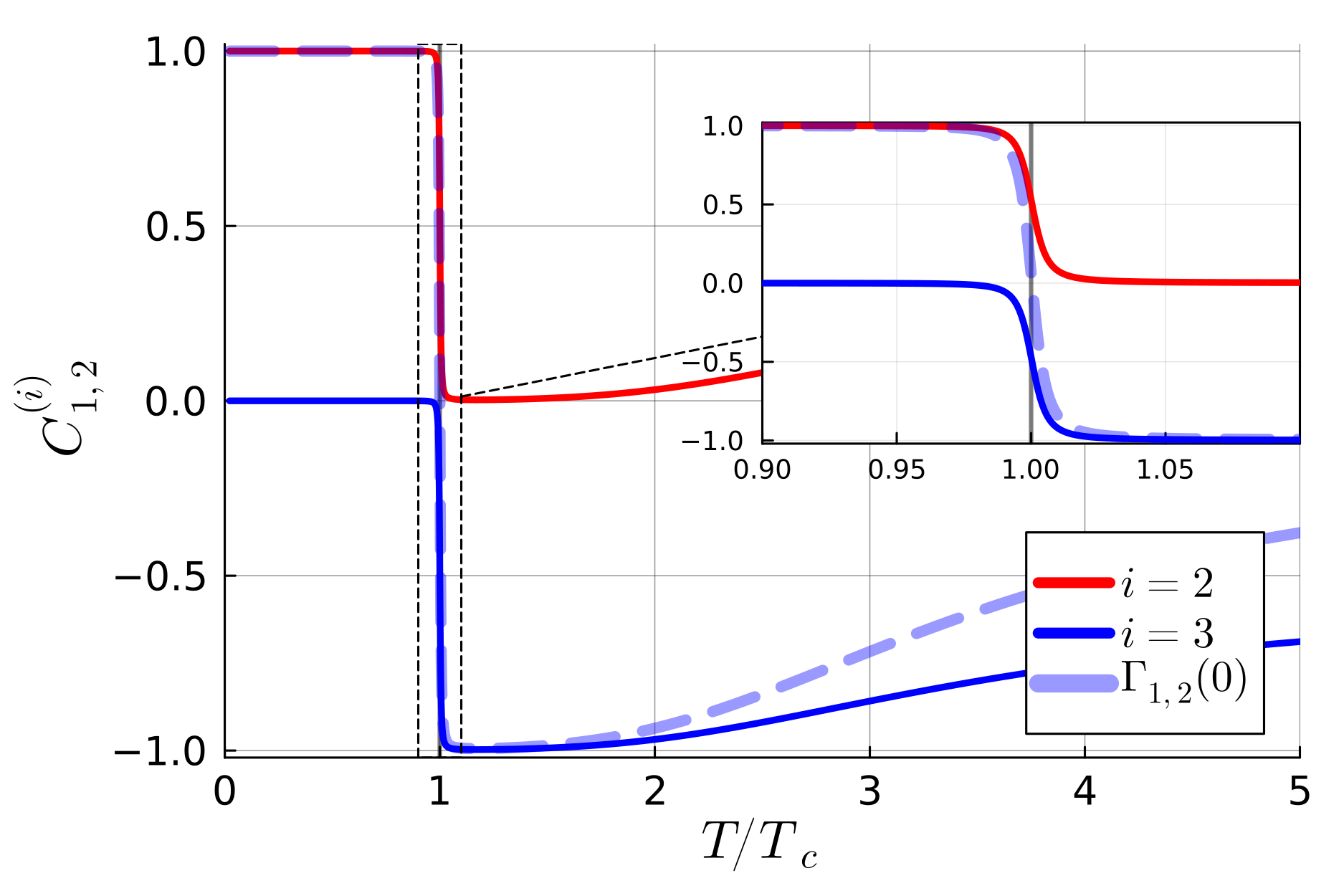}
		\caption{$C_{1,2}$}
		\label{fig:c12}
	\end{subfigure}
	\vfill
	\begin{subfigure}[b]{0.49\textwidth}
		\centering
		\includegraphics[width=\linewidth]{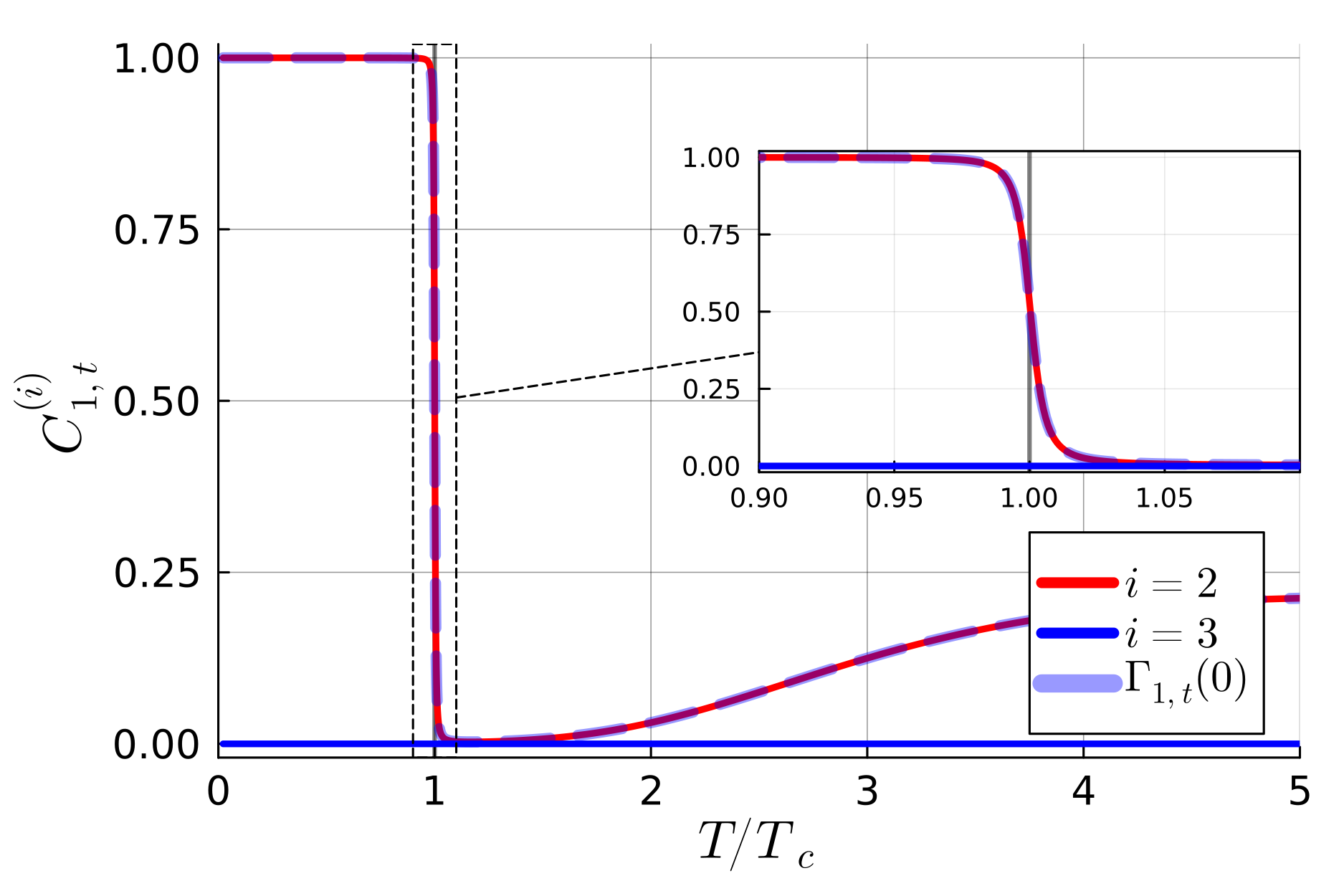}
		\caption{$C_{1,t}$}
		\label{fig:c1t}
	\end{subfigure}
	\hfill
	\begin{subfigure}[b]{0.49\textwidth}
		\centering
		\includegraphics[width=\linewidth]{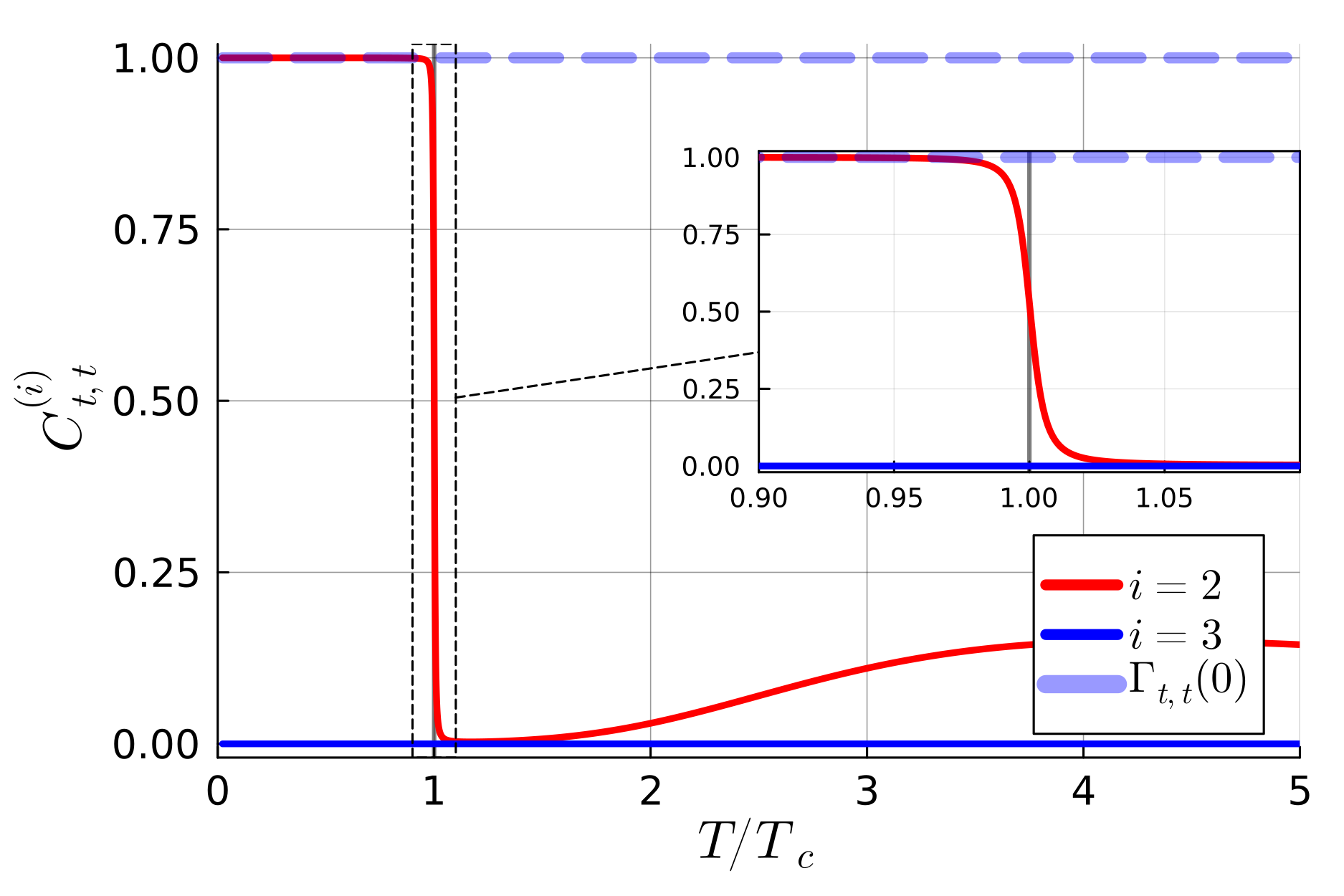}
		\caption{$C_{t,t}$}
		\label{fig:ctt}
	\end{subfigure}
	\caption{Plots of prefactors, with $J_{\mathrm{Leg}}=2, J_{\mathrm{Rung}}=-1,J_\Delta=1.2$. We include an additional curve (dashed light blue) to show the behaviour of $\Gamma_{\alpha,\beta}(0) = \sum_i C_{\alpha,\beta}^{(i)}$ in each plot. The inset shows the behaviour in the region around $T_c$, as calculated from \cref{tc_eq}. Note the near zero magnitude of the $i=2$ (shown in red) term in (c),(d) around $T_c/J_{\mathrm{Leg}} \approx 0.2885$. Note that in (d), we only show the $i=2,3$ terms. However, as indicated by the curve for $\Gamma_{t,t}(0)$, there are other prefactors that contribute for $R=0$.}
	\label{fig:prefactors}
\end{figure*}

There are four unique combinations of $\alpha,\beta$, where the rest of the combinations are equivalent by symmetry, these are
\begin{equation}\label{prefactor_combs}
	\alpha= 1,\ \beta = 1, \quad
	\alpha= 1,\ \beta = 2, \quad
	\alpha= 1,\ \beta = \mathrm{t},\quad
	\alpha= t,\ \beta = \mathrm{t}.
\end{equation}

Within the combinations in \cref{prefactor_combs}, we define two sets of prefactors. The first set relates to spins on the legs, $s_1,s_2$, and the other set involves at least one spin on the ``top'' (t). The behaviour of these prefactors is shown in \cref{fig:prefactors}, and shows the distinction between these two sets. The prefactors in \cref{fig:c11,fig:c12}, those relating to spins on the legs, behave in a distinct way to those in \cref{fig:c1t,fig:ctt}, relating to at least one spin on the ``top". The primary difference is that in \cref{fig:c1t,fig:ctt} we see the $i=3$ prefactor is \textit{exactly} zero, due to the matrix element $\braket{u_1|\mathbf{S}_t|u_3}$. We can also observe the behaviour discussed by W. Yin in \cref{fig:c12}, where $\Gamma_{1,2}$ changes sign \cite{yin2024paradigm}.

\begin{figure}[hbt]
	\centering
	\includegraphics[width=0.75\linewidth]{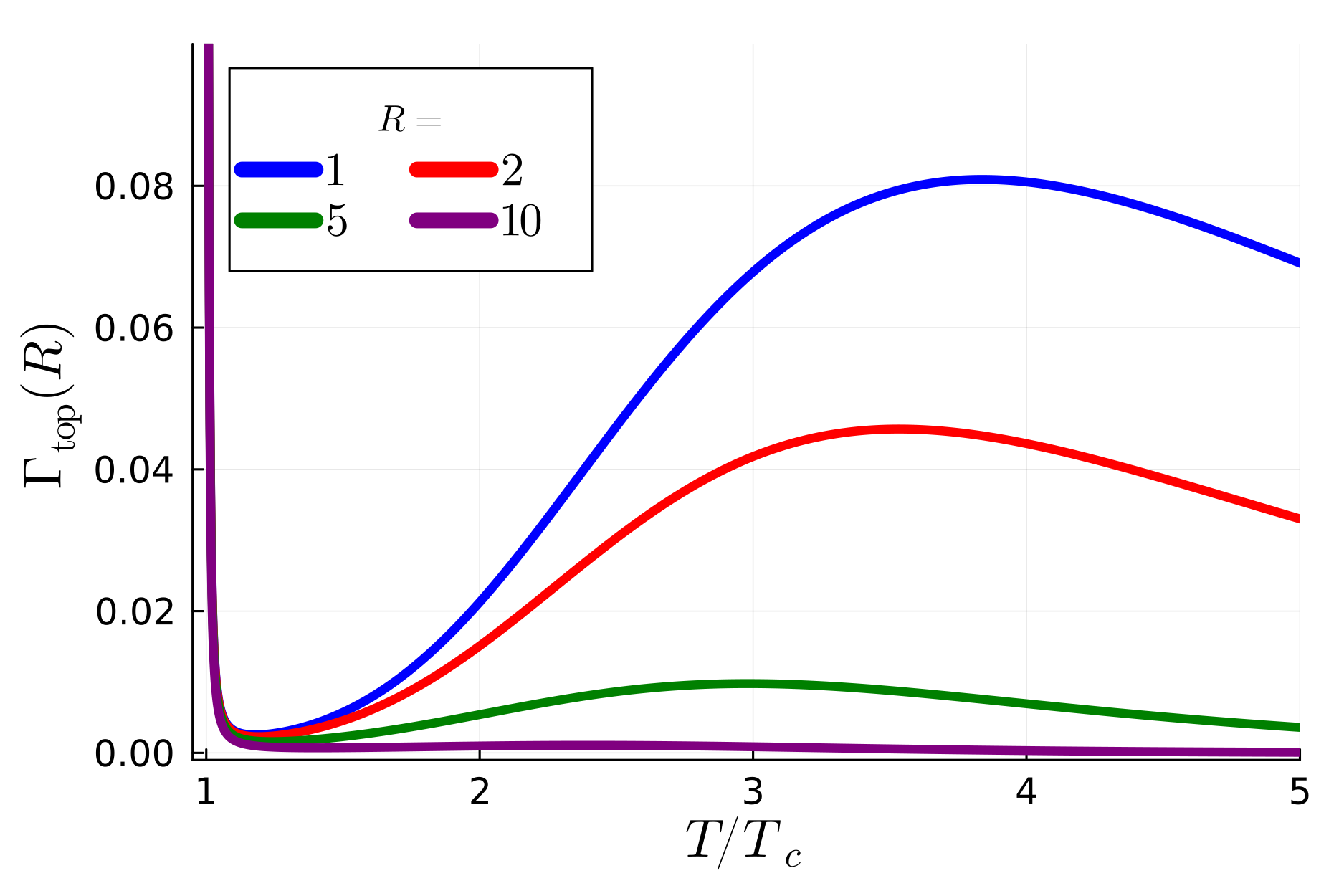}
	\caption{Plot of the correlation function between spins $s_{t,0},s_{t,R}$, for small values of $R$. We have used $J_{\mathrm{Leg}}=2, J_{\mathrm{Rung}}=-1,J_\Delta=1.2$. We can see that the non-monotonic behaviour, seen in the corresponding prefactor, persists for $R< 10$.}
	\label{fig:gamma_top}
\end{figure}

We see non-monotonic behaviour for the prefactors $C_{1,t},C_{t,t}$ in \cref{fig:c1t,fig:ctt}, and it is interesting to enquire if this will persist for non-zero $R$. This corresponds to the top spins becoming weakly coupled to each other, and the spins on the legs, in the vicinity of $T_c$, yet re-coupling upon increasing temperature. Clearly, $e^{-R/\xi}$ will dominate for large values of $R$, however we ask if there are some values of $R$ for which the non-monotonic behaviour will persist. We find that for small values of $R$, we can indeed observe $\Gamma_{\mathrm{top}}(R)$ displays the non-monotonic behaviour of the prefactor $C_{1,t}^{(2)}$. This is illustrated in \cref{fig:gamma_top} for $R = 1,2,5,10$. For $R<10$, the non-monotonic behaviour survives, but is dominated by the exponential term for $R\geq10$. This conveys that there is some short-range correlation between these top spins. 

We will now consider the consequence of $C_{\alpha,t}^{(3)} =0$, which can be related to the symmetries discussed in \cref{eigenvals_analytics}. The relevant eigenvector for this prefactor corresponds to the eigenvalue $\lambda_3$, which was separated in the $\hat{P_1},\hat{P_2} = -1,-1$ block. Thus the eigenvector must be anti-symmetric with respect to both $\hat{P_1}$, and $\hat{P_2}$, which causes the prefactor to be zero. 
The consequences of $C_{\alpha,t}^{(3)} =0$ can be seen when we take this prefactor in combination with the crossing of the eigenvalues $\lambda_2,\lambda_3$ in \cref{eigenvals_analytics}, and study the correlation length.

\subsection{The correlation length}\label{corr_len}

\begin{figure}[htb]
	\centering
	\includegraphics[width=0.75\linewidth]{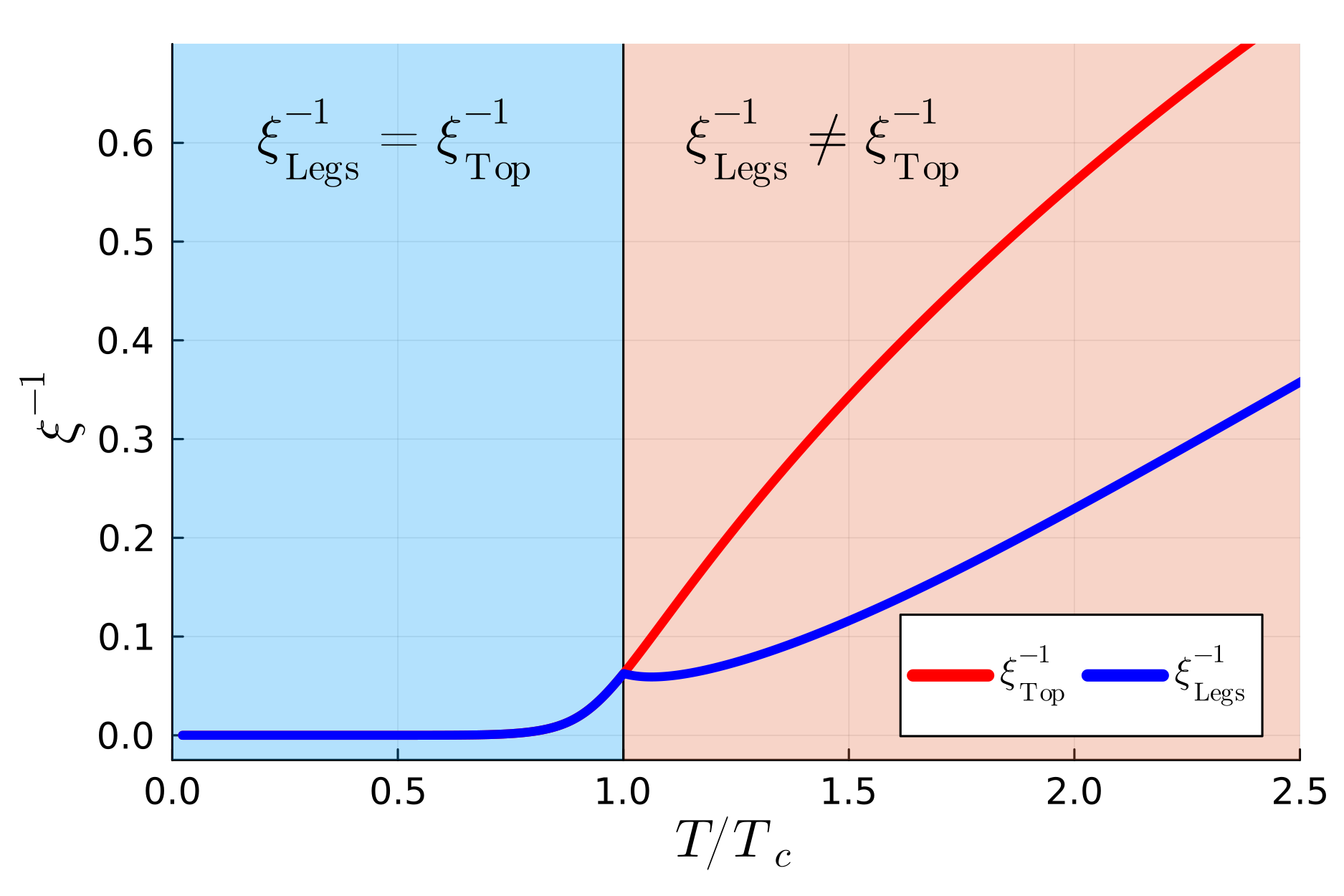}
	\caption{Plot showing the separation of the correlation lengths as a function of temperature. Here, we have used $J_{\mathrm{Leg}}=1, J_{\mathrm{Rung}}=-1,J_\Delta=1.2$. We see that at $T_c\approx 0.577$, there is a ``cusp" in the correlation length. This cusp signals the separation of the correlation lengths along the legs, $\xi^{-1}_{\mathrm{Legs}}$, and along the top, $\xi^{-1}_{\mathrm{Top}}$. This directly corresponds to the crossing of $\lambda_2,\lambda_3$.}
	\label{fig:phase_diagram_correlation_length}
\end{figure}

For ease of discussion, we restate the definition of the correlation function given in \cref{corr_defn}.

\begin{equation}
	\Gamma_{\alpha,\beta}\ (R) =  \sum_{i>1} e^{-R/\xi_i} \ C_{\alpha,\beta}^{(i)},
\end{equation}
where the correlation length is then
\begin{equation}
	\xi^{-1} =\lim_{R\to\infty} -\frac{1}{R}\ln \left[\sum_{i>1} \left(\frac{\lambda_i}{\lambda_1}\right)^R C_{\alpha,\beta}^{(i)}\right].
\end{equation}
In the limit $R\to\infty$, only the second largest eigenvalue will be responsible for the correlation length.  However, we saw that there is a crossing of $\lambda_2,\lambda_3$ in this model. Below the crossing temperature $T_c$, $\lambda_2$ describes the decay of correlations, whilst above $T_c$, $\lambda_3$ makes the significant contribution; the correlation length exhibits a discontinuity where these eigenvalues cross. This would be the whole story if not for the prefactors. We must take into account $C_{\alpha,t}^{(3)} =0$. The effect of this prefactor is that there is no contribution to the correlation function, and hence the correlation lengths, from the $i=3$ term for these top spins. These correlations are \textit{always} described by $\lambda_2$. This is not the case for the prefactors relating to the legs of the ladder, where the corresponding prefactors are non-zero. The correlations on the legs must then be described by $\lambda_3$ above $T_c$, whilst the correlations on the top are always described by $\lambda_2$. We have a \textit{bifurcation} of the correlation length at $T_c$; the relevant length scales for the legs and the top are distinct. Two regions naturally arise either side of this bifurcation. In \cref{fig:phase_diagram_correlation_length}, we clearly see the separation of the correlation lengths, and the formation of the two regions.

The correlation length between the top spins is shorter than the correlation length between the spins on the legs. This shorter length scale can be related to the frustration experienced by $s_{t,i}$ as was detailed in the preceding sections. A shorter correlation length means that that the ``order" between this spins decays more rapidly, which is in agreement with these spins being frustrated. Indeed, making this connection can start to probe the link between the crossing of $\lambda_{2},\lambda_3$ and the crossover seen in the thermodynamics. 

To understand this link, we consider the inclusion of an interaction between the spins $s_{t,i}$. Including such an interaction means that the top spins are no longer free to become frustrated, they instead act collectively due their mutual interaction. Recalling the mechanism for the crossover in the thermodynamics, we would now expect a reduction in the entropy released by these spins, and hence broadening the crossover. We define $J_\mathrm{Top}$ as the interaction between $s_{t,i}$ and $s_{t,i+1}$, details of this can be found in \ref{app_top}. Provided that $J_\mathrm{Top} < J_{\mathrm{Leg}}$, the transfer matrix can \textit{still} be block diagonalized in the same structure as discussed previously (\cref{eigenvals_analytics} and \ref{app_TM}), also when this interaction $J_\mathrm{Top}$ is included. Indeed, the two symmetries persist even when we include $J_{\mathrm{Top}}$. These symmetries mean that the criteria for the bifurcation in the correlation lengths are still met (the second and third largest eigenvalues can cross, and the corresponding prefactor can be exactly zero). However, as we see in \cref{fig:top_spec} of \ref{app_top}, the specific heat is broadened considerably by increasing $J_{\mathrm{Top}}$; the crossover is no longer ``ultra-narrow". The bifurcation in the correlation lengths, persists, and also becomes broader. We can understand that the shorter correlation length on the legs of the ladder, as $J_{\mathrm{Top}}$ is increased, causes a ``slower" gain of entropy, and thus a broader peak in the specific heat. 

The correlation length is a physically measurable quantity through the appropriate scattering experiments, which means this bifurcation could be measured. This does presuppose that a suitable candidate material exists, the discussion of which is outside the scope of the present work.

\section{Conclusions}

In this work, we have investigated the Ising model on the ``Toblerone" lattice (a decorated two-leg ladder, see \cref{tobl_Hamiltonian} and \cref{fig:tobl_sketch}) which has been previously shown to have a narrow peak in specific heat reminiscent of a phase transition in higher dimensional models. Through a careful study of magnetic susceptibilities, symmetries and eigenvalues of the transfer matrix, and correlation functions and lengths, we have elucidated the physical mechanisms driving this narrow peak.

In essence, the peak results from frustration in the model~\cite{yin2024paradigm}.  The ground state is unique — the two bottom legs have an effective ferromagnetic interaction between them due to interaction through the top spin.  This dominates over the direct antiferromagnetic interaction between the two bottom legs, $J_{\mathrm{Rung}}$.  However, there is then a manifold of degenerate states at low energy.  The physical picture is that when the temperature is sufficient to excite these states, the top spins become frustrated, and entropy is released, which leads to the peak in the specific heat. This effective interaction was explicitly studied in Ref.~\cite{hutak2021low}, where the top spin in each unit cell is integrated out, thus generating the effective interaction between the bottom two spins, $J_{\perp}(T)$. This interaction changes sign at $T_c$ from ferromagnetic, in the low temperature phase, to antiferromagnetic, in the high temperature phase.  When $J_\perp(T)$ is antiferromagnetic, the states have an extra degeneracy from the top spin — the degenerate manifold referred to above.  As the top spins are not directly connected to each other, they release an entropy of $\ln 2$ per unit cell.  The rate of this entropy released is governed by how collectively the bottom legs act — a large correlation length at $T_c$ leads to a narrow peak. Our results for the susceptibilities and correlation lengths back up this physical picture, and they assess subtleties that play an important role beyond its basic arguments.

In the low temperature phase, the ferromagnetic susceptibility dominates, as in an un-frustrated Ising model. In addition, the short distance correlation between the bottom legs is ferromagnetic, and at long distances the system is described by a single correlation length, as is the usual situation for one-dimensional classical models.

In contrast, in the high temperature phase we see a (sharp) crossover to a phase when the antiferromagnetic susceptibility dominates, and the short distance correlation between the bottom legs is now also antiferromagnetic.  More interestingly, we see a bifurcation in the correlation length at $T_c$ — in the high temperature phase the correlation length of the top spins is distinct and shorter than the correlation length of on the legs.  This shorter length scale for the top spins is an indicator of the frustration — although the spins have correlations through the base, they act more independently than those in the base.

Mathematically, this bifurcation in the correlation length is due to the crossing of the second and third eigenvalues of the transfer matrix, along with the fact that the prefactor for certain correlations is zero for $\lambda_3$, and the top spins. These both stem from certain symmetries of the model, discussed in detail in \cref{eigenvals_analytics}.  Interestingly, a direct coupling between the ``top'' spins do not break this symmetry and the bifurcation occurs all the way up to the isotropic case ($J_{\mathrm{Top}} = J_{\mathrm{Leg}}$). The bifurcation however becomes weaker (although still distinct, the two correlation lengths become closer together) along with the thermodynamic crossover becoming broader as this ``top'' coupling is increased.
With this, one can ask about the link between the two features. We have provided some analysis of this link in Appendix B, where we see that the bifurcation can persist when the thermodynamics are less sharp.  It is clear that these two features are arising from the same physics of frustration in the model, which is well understood. The exact mathematical link, however, remains an open question. 

This physical picture elucidates the details of the curious thermodynamics that were observed in the model of \cref{tobl_Hamiltonian}. Our work provides a more complete understanding about the mechanisms that can give rise to the narrow peak in the specific heat, especially through our results from the susceptibilities, the symmetries, and the correlation lengths.

\ack
JC was supported by the Engineering and Physical Sciences Research Council [EPSRC DTP 2022 EP/W52461X/1]. For initial discussions of this work the authors are thankful for the encouragement and hospitality by Tim Ziman at the Institute Laue-Langevin in Grenoble -- BT and JC acknowledge partial funding support by a QuantEmX grant from ICAM and the Gordon and Betty Moore Foundation through Grant GBMF5305. BT also acknowledges recent discussions with Mike Zhitomirsky and Tim Ziman which prompted deeper insights about the foundations of this work. SC thanks Luigi Amico and others at the Quantum Research Center in Abu Dhabi for their hospitality and useful discussions.

\appendix

\section{Transfer matrices: eigenvalues, symmetries, and spin matrices}\label{app_a}
\subsection{Solution of the model by Transfer matrices}\label{app_TM}
The transfer matrix for the model defined in \cref{tobl_Hamiltonian}, is:
\footnotesize
\begin{equation}\fl
\mathbf{T} = 
\bordermatrix{s_{1},s_{2},s_{t} & +,+,+ & +,+,- & +,-,+ & +,-,- & -,+,+ & -,+,- & -,-,+ & -,-,- \cr
	+,+,+ & x^{2} \; y \; z^{2} & x^{2} \; y & z & z & z & z & x^{-2}\; y  & x^{-2}\; y \; z^2 \cr
	+,+,- & x^{2} \; y & x^{2}\; y \; z^{-2}  & z^{-1} & z^{-1} & z^{-1} & z^{-1} &x^{-2}\; y\; z^{-2} & x^{-2}\; y  \cr
	+,-,+ & z & z^{-1} & x^{2} \; y^{-1} & x^{2} \; y^{-1} & x^{-2} \; y^{-1} & x^{-2} \; y^{-1} & z^{-1} & z \cr
	+,-,- & z & z^{-1} & x^{2} \; y^{-1} & x^{2} \; y^{-1} & x^{-2} \; y^{-1} & x^{-2} \; y^{-1} & z^{-1} & z \cr
	-,+,+ & z & z^{-1} & x^{-2} \; y^{-1} & x^{-2} \; y^{-1} & x^{2} \; y^{-1} & x^{2} \; y^{-1} & z^{-1} & z \cr
	-,+,- & z & z^{-1} & x^{-2} \; y^{-1} & x^{-2} \; y^{-1} & x^{2} \; y^{-1} & x^{2} \; y^{-1} & z^{-1} & z \cr
	-,-,+ & x^{-2}\; y  &x^{-2}\; y\; z^{-2} & z^{-1} & z^{-1} & z^{-1} & z^{-1} & x^{2}\; y \; z^{-2}  & x^{2} \; y \cr
	-,-,- & x^{-2}\; y \; z^2 & x^{-2}\; y  & z & z & z & z & x^{2} \; y & x^{2} \; y \; z^{2} \cr
},\label{transfer_mat}
\end{equation} 
\normalsize
where $x = \exp(J_{\mathrm{Leg}}/T), y = \exp(J_{\mathrm{Rung}}/T), z = \exp(J_\Delta/T)$. This matrix is as given by W. Yin~\cite{yin2024paradigm}. The basis for the transfer matrix is the spin states for $s_{1,i},s_{2,i},s_{t,i}$, and the neighbouring unit cell $s_{1,i+1},s_{2,i+1},s_{t,i+1}$. These are given above the rows and columns of the matrix respectively, where spin states of $s_{\alpha,i}=1$ are denoted by $+$, and $s_{\alpha,i}=-1$ are denoted by $-$.
The non-zero eigenvalues of the transfer matrix are
\begin{equation}
\lambda_1 =I_{+} \left[\cosh \frac{2J_{\mathrm{Leg}}}{T} + \sqrt{\frac{I_{-}^2}{I_{+}^2}\sinh^2 \frac{2J_{\mathrm{Leg}}}{T} + 1}\ \right]\label{lambda1_eq} 
\end{equation}
\begin{equation}
\lambda_2 = 4\  e^{J_{\mathrm{Rung}}/T} \sinh \frac{2J_{\mathrm{Leg}}}{T} \cosh\frac{2J_\Delta}{T}, 		\label{lambda2_eq} 
\end{equation}
\begin{equation}
\lambda_3 = 4\ e^{-J_{\mathrm{Rung}}/T} \sinh \frac{2J_{\mathrm{Leg}}}{T},												\label{lambda3_eq}
\end{equation}
\begin{equation}
\lambda_4 =I_{+} \left[\cosh \frac{2J_{\mathrm{Leg}}}{T} - \sqrt{\frac{I_{-}^2}{I_{+}^2}\sinh^2 \frac{2J_{\mathrm{Leg}}}{T} + 1}\ \right],	\label{lambda4_eq}
\end{equation}

with $I_{\pm} = 2e^{J_{\mathrm{Rung}}/T}\cosh \frac{2J_\Delta}{T} \pm 2e^{-J_{\mathrm{Rung}}/T}$. These coincide with the so called ``frustration functions", $\Upsilon_\pm$  in Ref.~\cite{yin2024paradigm}. 

\subsection{Symmetries of the transfer matrix}\label{app_sym}
There are two symmetries of the transfer matrix in \cref{transfer_mat} that we focus on here. The spin-flip, $\hat{P_1}$, and the leg-exchange, $\hat{P_2}$, symmetries are given by the following operators

\footnotesize
\begin{equation}\hspace{-3em}
\hat{P}_1 = \left[
\begin{array}{cccccccc}
	0 & 0 & 0 & 0 & 0 & 0 & 0 & 1 \\
	0 & 0 & 0 & 0 & 0 & 0 & 1 & 0 \\
	0 & 0 & 0 & 0 & 0 & 1 & 0 & 0 \\
	0 & 0 & 0 & 0 & 1 & 0 & 0 & 0 \\
	0 & 0 & 0 & 1 & 0 & 0 & 0 & 0 \\
	0 & 0 & 1 & 0 & 0 & 0 & 0 & 0 \\
	0 & 1 & 0 & 0 & 0 & 0 & 0 & 0 \\
	1 & 0 & 0 & 0 & 0 & 0 & 0 & 0 \\
\end{array}
\right],\ 
\hat{P}_2 = \left[
\begin{array}{cccccccc}
	1 & 0 & 0 & 0 & 0 & 0 & 0 & 0 \\
	0 & 1 & 0 & 0 & 0 & 0 & 0 & 0 \\
	0 & 0 & 0 & 0 & 1 & 0 & 0 & 0 \\
	0 & 0 & 0 & 0 & 0 & 1 & 0 & 0 \\
	0 & 0 & 1 & 0 & 0 & 0 & 0 & 0 \\
	0 & 0 & 0 & 1 & 0 & 0 & 0 & 0 \\
	0 & 0 & 0 & 0 & 0 & 0 & 1 & 0 \\
	0 & 0 & 0 & 0 & 0 & 0 & 0 & 1 \\
\end{array}
\right].
\end{equation}
\normalsize
As discussed in the main text, these symmetries commute with both the transfer matrix and with each other, allowing a simultaneous diagonalisation with a single unitary matrix $U$. We construct this matrix from the eigenstates (and linear combinations) of the symmetry operators $\hat{P_1},\hat{P_2}$. This matrix is

\footnotesize
\begin{equation}
\centering
U = \left[
\begin{array}{cccccccc}
	\frac{1}{\sqrt{2}} & 0 & 0 & 0 & 0 & 0 & 0 & \frac{1}{\sqrt{2}} \\
	0 & \frac{1}{\sqrt{2}} & 0 & 0 & 0 & 0 & \frac{1}{\sqrt{2}} & 0 \\
	0 & 0 & \frac{1}{2} & \frac{1}{2} & \frac{1}{2} & \frac{1}{2} & 0 & 0 \\
	0 & 0 & \frac{1}{2} & -\frac{1}{2} & \frac{1}{2} & -\frac{1}{2} & 0 & 0 \\
	0 & 0 & \frac{1}{2} & -\frac{1}{2} & -\frac{1}{2} & \frac{1}{2} & 0 & 0 \\
	0 & 0 & \frac{1}{2} & \frac{1}{2} & -\frac{1}{2} & -\frac{1}{2} & 0 & 0 \\
	0 & \frac{1}{\sqrt{2}} & 0 & 0 & 0 & 0 & -\frac{1}{\sqrt{2}} & 0 \\
	\frac{1}{\sqrt{2}} & 0 & 0 & 0 & 0 & 0 & 0 & -\frac{1}{\sqrt{2}} \\
\end{array}\right].
\end{equation}
\normalsize
By then taking the product $U^T\ \mathbf{T}\ U$, a block diagonal transfer matrix is obtained

\begin{equation}
U^T\ \mathbf{T}\ U =\left[\begin{array}{cccc}\mathbf{A}_1 & 0 & 0&0\\ 	0& \mathbf{A}_2  & 0&0\\0 & 0& \mathbf{A}_3 & 0\\ 0 & 0 &0 & \mathbf{A}_4 \end{array} \right],
\end{equation}
where the blocks $\mathbf{A}_i$ correspond to the eigenvalues of $\hat{P_i}$, and are defined as follows

\begin{eqnarray}\fl
		{P_1} = +1, {P_2} =+1 &: \
		\mathbf{A}_1 = \left[
		\begin{array}{ccc}
			x^{2} \  y \  z^{2} + \frac{  y \  z^{2}}{x^{2}} &   x^{2} \  y + \frac{  y}{x^{2}} &  2 \sqrt{2} \  z \\
			x^{2} \  y + \frac{  y}{x^{2}} & \frac{  x^{2} \  y}{z^{2}} + \frac{  y}{x^{2} \  z^{2}} & \frac{ 2 	\sqrt{2}}{z} \\
			2 \sqrt{2} \  z & \frac{ 2 \sqrt{2}}{z} & \frac{2 \  x^{2}}{y} + \frac{2}{x^{2} \  y} \\
		\end{array}\right],  \\[1em]
		\fl {P_1} = +1, {P_2} = -1 &: \	\mathbf{A}_2 =  0, \\[1em]
		\fl {P_1} = -1, {P_2} = -1 &: \ \mathbf{A}_3 =  \frac{2 \  x^{2}}{y} - \frac{2}{x^{2} \  y}, \\[1em]
		\fl	{P_1} = -1, {P_2} = +1 &: \	\mathbf{A}_4 =  \left[\begin{array}{ccc}
			0 & 0 & 0 \\
			0 & \frac{  x^{2} \  y}{z^{2}} - \frac{  y}{x^{2} \  z^{2}} &   x^{2} \  y - \frac{  y}{x^{2}} \\
			0 &   x^{2} \  y - \frac{  y}{x^{2}} &   x^{2} \  y \  z^{2} - \frac{  y \  z^{2}}{x^{2}} \\
		\end{array}\right].  
\end{eqnarray}
The eigenvalues $\lambda_1, \lambda_4$, \cref{lambda1_eq,lambda4_eq} can be found from the block $\mathbf{A}_1$, $\lambda_{2}$,\cref{lambda2_eq}, from block $\mathbf{A}_4$, and $\lambda_3$, \cref{lambda3_eq} from block $\mathbf{A}_3$.

\subsection{Spin Matrices}\label{app_spin}

Computing the correlation functions for the model requires use of the appropriate diagonal matrices, $\mathbf{S}_{\alpha,\beta}$. The elements of $\mathbf{S}_{\alpha,\beta}$ can be deduced from the basis of the transfer matrix, shown in \cref{transfer_mat}. For the three spins in each unit cell, the diagonals of the corresponding matrices are

\begin{equation}
\mathbf{S}_{1} = \left[\begin{array}{cccccccc}
	1 & 1 & 1 & 1 & -1 & -1 & -1 & -1 \\
\end{array}\right],
\end{equation}
\begin{equation}
\mathbf{S}_{2} = \left[\begin{array}{cccccccc}
	1 & 1 & -1 & -1 & 1 & 1 & -1 & -1 \\\end{array}\right],
\end{equation}
\begin{equation}
\mathbf{S}_{\mathrm{t}} = \left[\begin{array}{cccccccc}
	1 & -1 & 1 & -1 & 1 & -1 & 1 & -1 \\
\end{array}\right].
\end{equation}

\section{The effect of including a coupling between the ``top" spins}\label{app_top}

\begin{figure}[htb]
\centering
\includegraphics[width=0.5\textwidth]{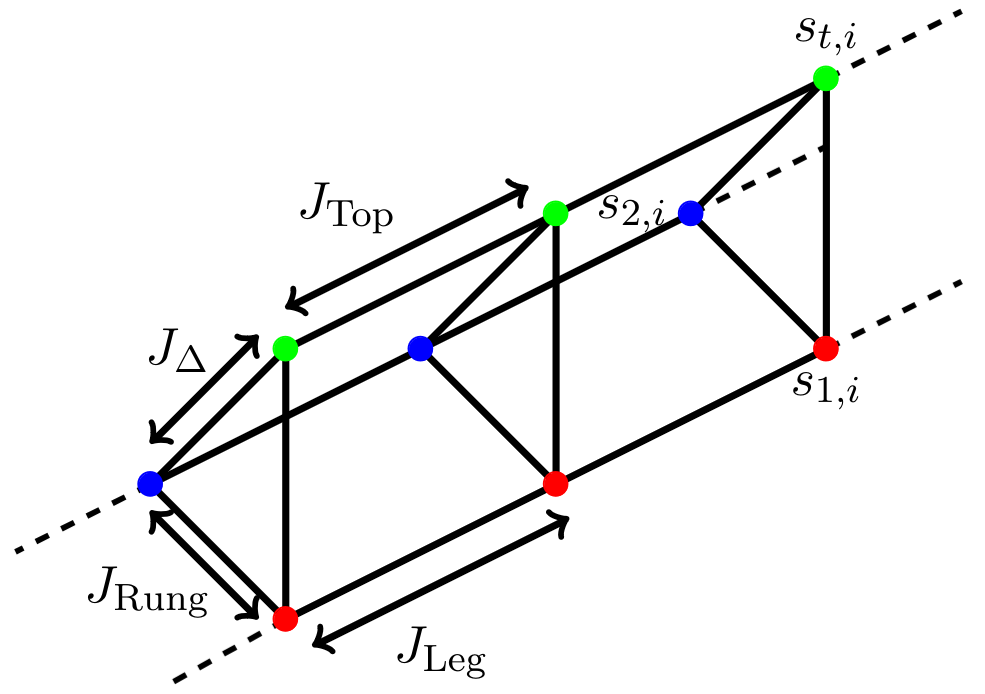}    
\caption{Sketch of the ``Toblerone" lattice with the inclusion of the coupling between $s_{t,i}$ through $J_\mathrm{Top}$, as in \cref{top_j_eq}. The rest of the lattice remains unchanged from \cref{fig:tobl_sketch} and \cref{tobl_Hamiltonian}.}
\label{fig:tobl_top_sketch}
\end{figure}

It is instructive to consider the effect of including an interaction between the spins $s_{t,i}, s_{t,i+1}$. We include the additional term in the Hamiltonian
\begin{equation}\label{top_j_eq}
\mathcal{H}_{\mathrm{top}} = - J_{\mathrm{Top}}\sum_i s_{t,i} s_{t,i+1},
\end{equation}
where $J_\mathrm{Top}$ mediates the interaction between the spins $s_{t,i}$ and $s_{t,i+1}$. The case of $J_{\mathrm{Top}}= J_\mathrm{Leg}$ corresponds to the three-leg ladder with boundary conditions across the rungs, therefore we focus on $J_{\mathrm{Top}} < J_{\mathrm{Leg}}$ \cite{mejdani1996vII}. We sketch this in \cref{fig:tobl_top_sketch}.

The unitary matrix $U$ is found to also block diagonalise the transfer matrix corresponding to the inclusion of the top coupling. The resulting block diagonal matrix has the same structure as in the absence of the additional interaction, but the blocks themselves are distinct. We no longer find four eigenvalues are zero, which relates to the inability to trace out the top spins when we include the additional interaction. 

\begin{figure}[!htb]
\begin{subfigure}[c]{0.5\linewidth}
	\centering
	\includegraphics[width=\linewidth]{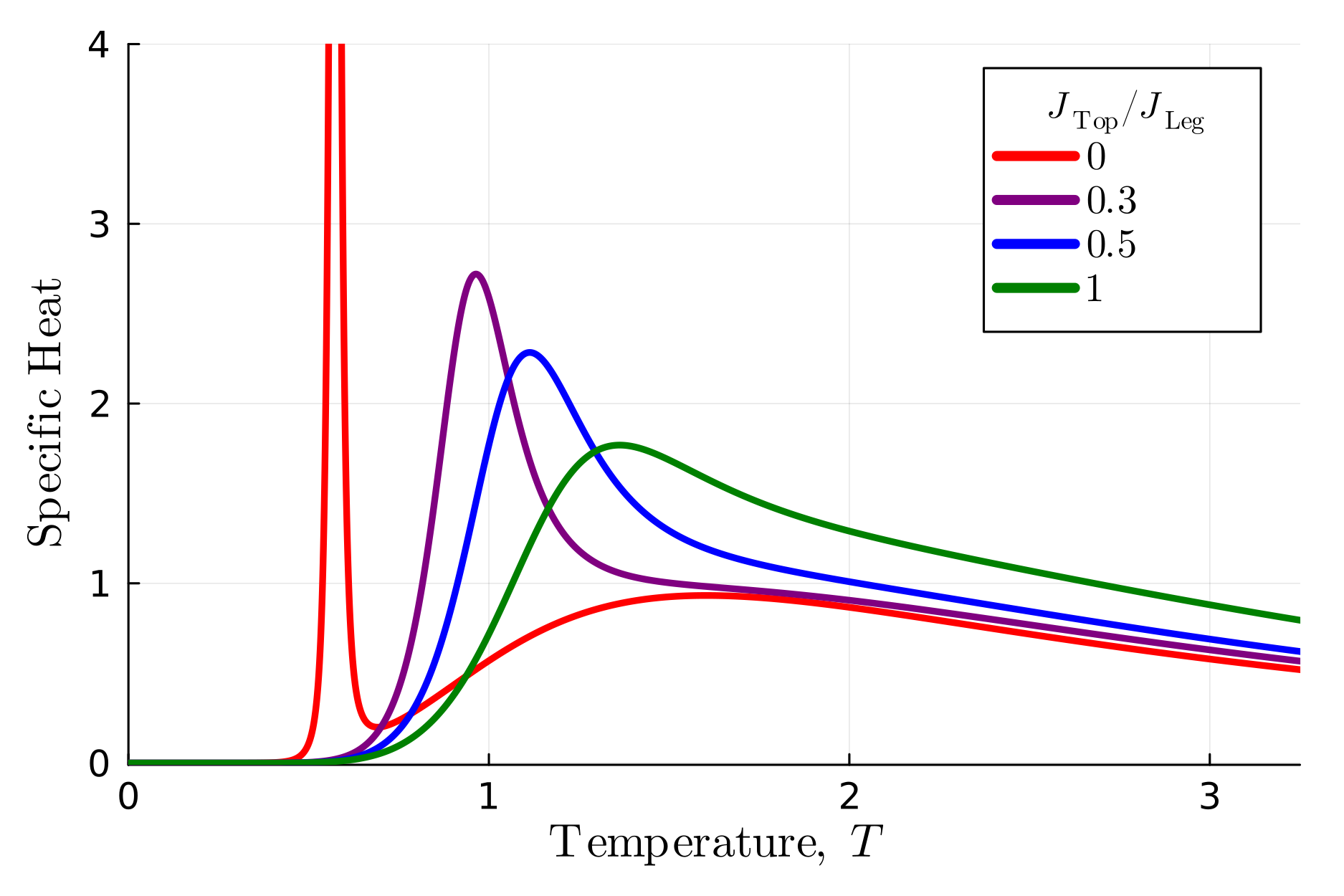}
	\caption{}
	\label{fig:top_spec}
\end{subfigure}
\begin{subfigure}[c]{0.5\linewidth}
	\centering
	\includegraphics[width=\linewidth]{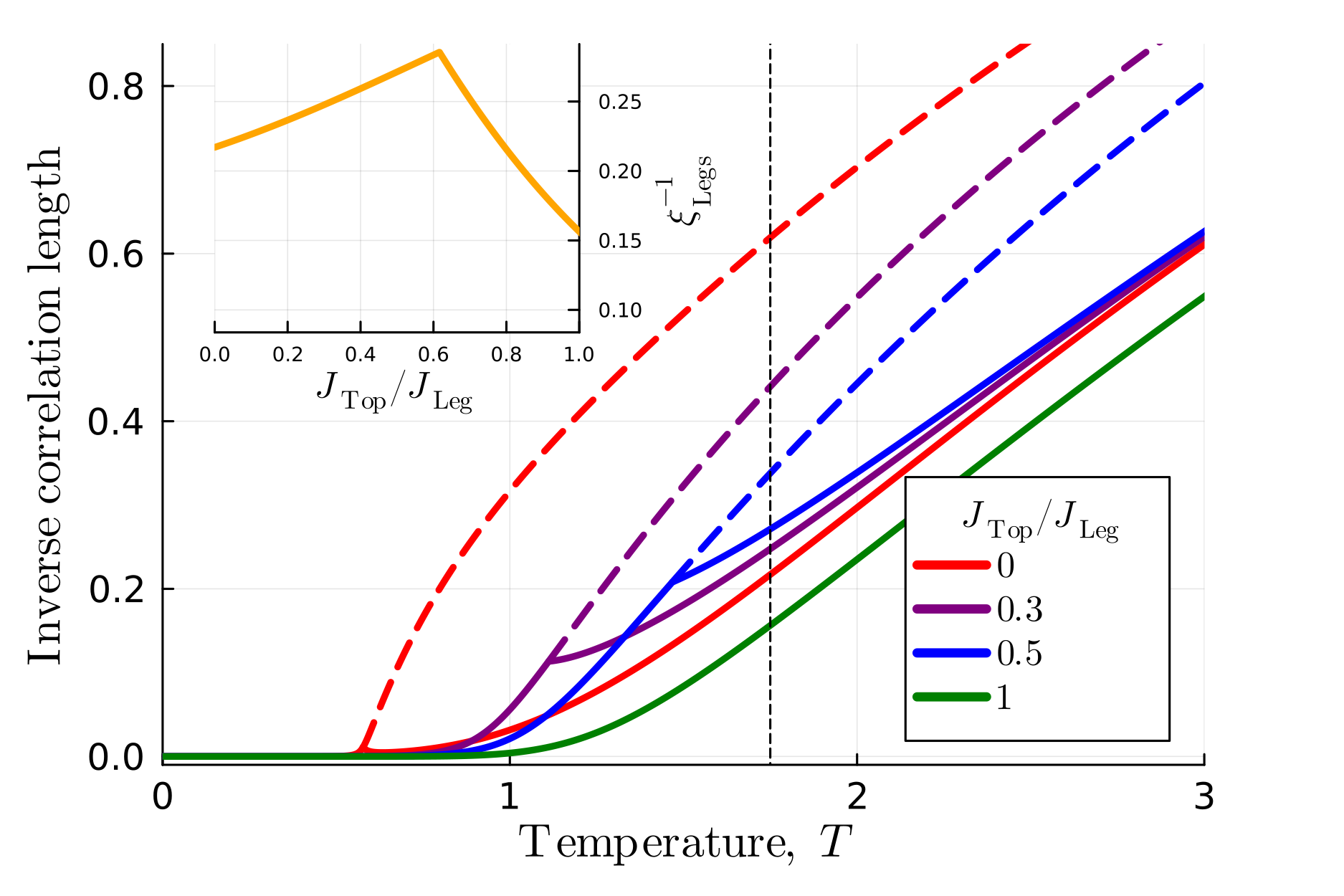}
	\caption{}
	\label{fig:top_xi}
\end{subfigure}
\label{top_plots}
\caption{Plots of the specific heat [panel (a)], and inverse correlation lengths [panel (b)] for $J_{\mathrm{Top}}/J_{\mathrm{Leg}}~=~0~,~0.3~,~0.5~,~1$. Panel (a): observe the peak in the specific heat broadening. We use $J_{\mathrm{Leg}}=1.5, J_{\mathrm{Rung}}=-1,J_\Delta=1.2$ for this plot. For the case of $J_{\mathrm{Top}}=0$,the peak in the specific heat extends past the top of the plot, but remains of finite height. Panel (b): The solid lines indicate the correlation length on the legs, and the dashed lines correspond to the top spins. The inset of panel (b) shows the inverse correlation length on the legs as a function of $J_{\mathrm{Top}}/J_{\mathrm{Leg}}$, at an arbitrarily chosen fixed temperature, $T=1.75$. This temperature is shown by the thin vertical line.}
\end{figure}

By considering the nature of the crossover described in the main text, the rapid increase in entropy on account of the top spins becoming frustrated, we can question if the inclusion of $J_\mathrm{Top}$ will have any effect on the thermodynamics. Indeed, we see that this coupling results in these spins are now less ``free" to be become frustrated. This behaviour is reflected in \cref{fig:top_spec}, where increasing values of $J_{\mathrm{Top}}$ can be seen to broaden the behaviour of the specific heat; the crossover is no longer ``ultra-narrow". Indeed, this broadening of the peak also corresponds to a broadening of the bifurcation in the correlation lengths, as shown in \cref{fig:top_xi}.

The bifurcation persists for $-|J_\Delta + J_{\mathrm{Rung}}| < J_{\mathrm{Top}} < J_\mathrm{Leg}$, but occurs at higher temperatures with increasing $J_{\mathrm{Top}}$. This is shown in \cref{fig:top_xi}. We observe that increasing $J_{\mathrm{Top}}$ results in the inverse correlation length \textit{decreasing} between the top spins (shown in the dashed lines). This corresponds to an \textit{increase} in the correlation length between the top spins, thus the interaction is acting to reduce the frustration of these spins.
Consider now, the behaviour of the correlation length on the legs. We observe non-monotonic behaviour as a function of $J_{\mathrm{Top}}$. This can be seen in the inset of \cref{fig:top_xi}, where we plot $\xi^{-1}_{\mathrm{Legs}}$ as a function of $J_{\mathrm{Top}}/ J_{\mathrm{Leg}}$ for an fixed temperature above the bifurcation point. We see that the inverse correlation length exhibits a "kink" for some value of $J_{\mathrm{Top}}$. This "kink" provides the value of $J_\mathrm{Top}/J_{\mathrm{Leg}}$ at which the correlation length will undergo a bifurcation at a given temperature. The initial decrease, with increasing $J_{\mathrm{Top}}$, in correlation length on the legs is interpreted as the frustration being ``shared" to the legs as the top spins are less easily frustrated. This frustration is reduced for further increasing $J_{\mathrm{Top}}$, leading to the increase in correlation length on the legs.

\vspace{5em}

\bibliographystyle{iopart-num}
\bibliography{main}

\end{document}